\documentclass{article}
\pdfoutput=1

\usepackage{graphicx}

\usepackage[english]{babel}

\usepackage[letterpaper,top=2cm,bottom=2cm,left=3cm,right=3cm,marginparwidth=1.75cm]{geometry}

\usepackage{amsmath}
\usepackage{graphicx}
\usepackage[colorlinks=true, allcolors=blue]{hyperref}

\title{\textbf{Pedagogical Implications for the Falling Astronaut Problem}}
\author{\textbf{Scott C. Scharlach$^1$}}

\begin{document}

\maketitle

\begin{center}
{\textit{$^1$Tufts University \\
574 Boston Avenue \\
Medford, MA 02155, USA}}
\end{center}

\begin{abstract}
 This paper outlines a deceptively complex problem in classical mechanics which the paper names the ``Falling Astronaut Problem,'' and it explores a method for teachers to implement this problem in an undergraduate classroom. The paper presents both an analytical solution and a numerical approximation to the Falling Astronaut Problem and compares the educational merit of the two approaches. The analytical solution is exact; however, the derivation requires techniques that are more advanced than what is typically seen in an introductory undergraduate physics course. In contrast, the numerical approximation presents a novel application of concepts with which a first-semester undergraduate is likely to be familiar. The paper stresses the pedagogical implications of this problem, specifically the opportunity for introductory undergraduate students to learn the utility of differential equations, numerical approximations, and data spreadsheets. On a more fundamental level, the paper argues that the Falling Astronaut Problem presents an instructional opportunity for physics students to acquire a nuanced and informative lens through which to conceptualize cause and effect in the universe.

\end{abstract}

\begin{center}
{\textit{Key Words}: Physics Education Research -- Differential Equations -- Numerical Approximation -- Newtonian Gravity}
\end{center}

\section{Introduction}

Differential equations are of critical importance in physics. They are fundamental to quantum mechanics via the Schrodinger Equation and to General Relativity via the Einstein field equations, and they have numerous applications to engineering and machinery via the simple harmonic oscillator equation. However, despite the essential role that differential equations play in modeling the universe, as well as their ubiquity across various research fields, introductory undergraduate physics classes tend to refrain from emphasizing differential equations. Indeed, some courses intended for first-semester college students may intentionally avoid differential equations when possible. Some physics curricula appear to hold an implicit assumption that differential equations are inappropriately advanced for a first-semester college student, and that differential equations ought to be reserved for upper-division courses.

This paper argues that some differential equations are not only appropriate for first-semester undergraduates, but in fact that differential equations ought to be embraced as an efficient and insightful tool for modeling systems, especially in classical mechanics.

To demonstrate the utility of differential equations, the paper presents a physics problem concerning Newtonian gravity that can be solved through both analytical and numerical techniques. The paper argues that the numerical approximation using differential equations relies only on mathematical concepts that are often taught in high school calculus courses, and which first-semester college students are likely to be familiar with, although the mathematics are applied in a new and instructive manner. The author believes that introducing students to differential equations early in their college curriculum will help students develop a deep and nuanced schema for conceptualizing motion and rates of change.

\section{Problem Statement}
\label{ProblemStatement}

An object such as an astronaut is released from a height $h$ above a spherical planet with mass $M$. The astronaut has negligible mass and volume compared to the planet. The astronaut initially has no angular momentum or linear momentum. The astronaut falls without air resistance until it reaches the surface of the planet at a height $R$.

\begin{figure}[ht]
\centering
\includegraphics[width=0.2\textwidth]{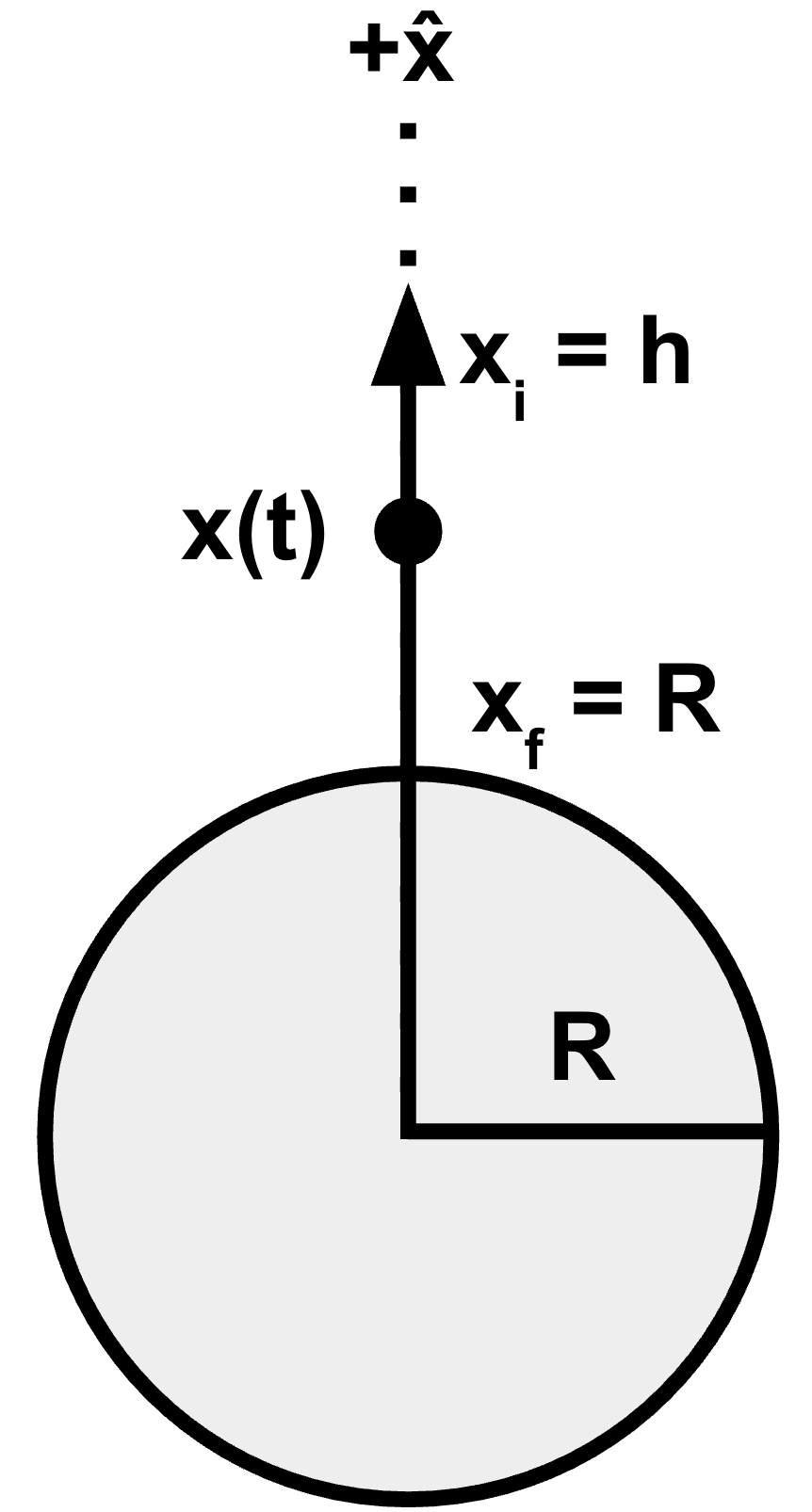}
\caption{The position of the astronaut at an arbitrary time $t$, which we call $x(t)$ or simply $x$, is located between the initial position $x_i = h $ and the final position $x_f = R$. Note that the final position is equal to the radius of the planet, as the astronaut lands on the planet's surface. The coordinate system defines positive $x$ as being directed away from the center of the planet.}
\label{diagram1}
\end{figure}

The planet has a mass of $5.97219 \times 10^{24}$ kg and a uniform radius of $6.371 \times 10^{6} $ m. In other words, the planet resembles Earth but without an atmosphere and with a perfectly spherical shape. The universal gravitational constant $G$ is $6.6743 \times 10^{-11} $ m$^{3}$ kg$^{-1}$ s$^{-2}$. The astronaut is dropped from a height of $1.0000 \times 10^{7}$ m. The astronaut is released at time $t_i$ = 0 s and lands on the ground at time $t_f$. A diagram visualizing the astronaut and the planet is shown in Figure \ref{diagram1}.

Note that the acceleration due to gravity $g$ is not constant as the astronaut approaches the planet's surface; $g$ only reaches the familiar value of 9.8 m/s$^2$ when the object is very close to the surface of the Earth.

\textit{(a)} Find $t_f$ in seconds. In other words, how much time passes between the astronaut being released and the astronaut reaching the surface? 

\text{(b)} What is the height $x$ for any given time $t$?

\section{Analytical Solution}

The following two sections are written in a pedagogical style intended to be approachable for first-semester undergraduate students. Any deviation from a traditional academic tone is intentional for educational purposes.

\subsection{Summary of the Technique}

The Falling Astronaut Problem may initially appear straightforward. We may be tempted use a one-dimensional kinematic equation such as

\begin{equation}
    \label{Kinematic}
    \Delta x = -\frac{1}{2} g \Delta t^2 +v_i \Delta t
\end{equation}

and solve for $\Delta t$. Such an approach implicitly assumes the acceleration due to gravity $g$ to be a constant. However, the initial height $h$ is large enough that the astronaut will experience an acceleration substantially less than what is experienced on the surface of the Earth. For this reason, we cannot use any of the constant-acceleration kinematic equations, including Equation \ref{Kinematic}.

The analytical approach presented in this section begins with the principle of conservation of energy. We then separate time and position so that the left-hand side of the equation contains time $t$ but not position $x$, while the right-hand side contains $x$ but not $t$. Both sides are then integrated. The time-dependent integral is straightforward. However, the position-dependent integral requires a u-substitution, an identity using functions called the ``Beta" and ``Gamma" functions, and two trigonometric identities.

For clarity, we restate our variables and assumptions here:

\begin{itemize}
    \item $h \equiv$ initial altitude relative to center of Earth (constant)
    \item $R \equiv$ radius of Earth (constant)
    \item $x \equiv$ current position at time $t$ (variable, function of $t$)
    \item $M \equiv$ mass of planet (constant)
    \item $m \equiv$ mass of astronaut (constant)
    \item The object is initially at rest, $x'(t=0s) = 0$ m/s
    \item There is no air resistance or rotation.
    \item \textbf{Goal:} Find $t$ for any given $x$, especially when $x = R$.
\end{itemize}

\subsection{Conservation of Energy}

The only two relevant forms of energy for the falling astronaut are kinetic and potential, thus conservation of energy gives us:

\begin{equation}
\label{eq:energy}
   E_i = E_t \Rightarrow K_i + U_i = K_t + U_t .
\end{equation}

The speed is non-relativistic, so we are justified in using the classical mechanics equation

\begin{equation}
\label{eq:kinetic}
   K = \frac{1}{2}mv^2 = \frac{1}{2}m \left( \frac{dx}{dt} \right)^2.
\end{equation}

The acceleration due to gravity $g$ is not constant, so we must use the equation

\begin{equation}
\label{eq:potential}
   U = - \frac{G M m}{x}
\end{equation}

instead of the more compact

\begin{equation}
\label{eq:potential2}
   U = m g x.
\end{equation}

The object is initially at rest ($\frac{dx}{dt}=0$ m s$^{-1}$) and at a height $x_i=h$, so

\begin{equation}
    E_i = K_i + U_i = 0 -\frac{G M m}{h} = -\frac{G M m}{h} .
\end{equation}

At an arbitrary time $t$, the energy is

\begin{equation}
\label{eq:E(t)}
    E_t = K_t + U_t = \frac{1}{2} m \left( \frac{dx}{dt} \right)^2 -\frac{G M m}{x}.
\end{equation}

Thus, $E_i = E_t$ becomes

\begin{equation}
\label{EiEF}
-\frac{G M m}{h} = \frac{1}{2} m \left( \frac{dx}{dt} \right)^2 - \frac{G M m}{x}.
\end{equation}

Dividing both sides by $m$,

\begin{equation}
\label{nomp}
-\frac{G M}{h} = \frac{1}{2} \left( \frac{dx}{dt} \right)^2 - \frac{G M}{x} .
\end{equation}

Note that the mass of the astronaut plays no role in its motion as she falls. In other words, two astronauts of different masses will have identical motion in this problem.

We now have a succinct equation relating all of our relevant variables. However, we do not yet have time as a function of position -- we need to separate position and time.

\subsection{Separation of Variables}

Equation \ref{nomp} has both the variable $x$ and the first derivative of that variable, $\frac{dx}{dt}$. An equation in which a variable is defined in terms of its own derivative is called a ``differential equation." Although this equation depends on time, as seen in the derivative operator $\frac{d}{dt}$, the variable $t$ does not explicitly appear. (The derivative $\frac{d}{dt}$ is an operator, not a variable.) Thus, we have not yet solved the problem because we cannot directly insert a particular time (such as $t = 100 s$) and calculate the position $x$ at that time.

The next step is to find an ``analytical" form for this differential equation. A function is analytical if it is exact, has a finite number of terms, accepts in an independent variable and outputs a dependent variable, and relies on standard mathematical operators such as multiplication, trigonometric functions, logarithms, et cetera. (In layman's terms, an analytical equation is a normal equation, in contrast to a differential equation.)

It is sometimes possible to convert a differential equation into an analytical equation -- such an equation is called an ``analytical solution." To find an analytical solution to this differential equation, we will take Equation \ref{nomp} and rewrite it so that the left-hand side of the equation references only time and constants, but not position, and the right-hand side of the equation references only position and constants, but not time:

\begin{equation}
\label{separated}
    \begin{gathered}
        -\frac{G M}{h} = \frac{1}{2} \left( \frac{dx}{dt} \right)^2 - \frac{G M}{x} \\[10pt]
        \Rightarrow \frac{G M}{x} - \frac{G M}{h} = \frac{1}{2} \left( \frac{dx}{dt} \right)^2 \\[10pt]
        \Rightarrow 2 G M \left( \frac{1}{x} - \frac{1}{h} \right) = \left( \frac{dx}{dt} \right)^2 \\[10pt]
        \Rightarrow \pm \sqrt{2 G M} \sqrt{\frac{1}{x} - \frac{1}{h}} = \frac{dx}{dt}
    \end{gathered}
\end{equation}

We are close to separating position and time, but first let us inspect Equation \ref{separated}, which tells us the velocity at any given moment in time.

The square root operation leaves an ambiguity: the velocity might be positive or negative. By inspecting Figure \ref{diagram1}, we see that positive vectors are pointed away from the center of the planet, while negative vectors are pointed toward the center of the planet. Our astronaut falls toward the planet, thus our velocity vector must be negative. For that reason, we proceed with only the negative value for Equation \ref{separated}.

We now separate our variables by multiplying both sides by $dt$ and dividing both sides by $\sqrt{\frac{1}{x}-\frac{1}{h}}$:

\begin{equation}
\label{separated2}
-\sqrt{2 G M} \, dt = \frac{1}{\sqrt{\frac{1}{x} - \frac{1}{h}}} \, dx .
\end{equation}

Are we justified in multiplying both sides by $dt$? As stated earlier, $t$ is not a variable, but rather it is one of the symbols composing the derivative operator $\frac{d}{dt}$, which can be thought of as a single mathematical concept but written with more than one symbol.

Physicists are indeed justified in making this mathematical step. Taken literally, $\frac{d}{dt}x$ means ``the derivative operator applied to the position at a given instant in time." However, such a concept is approximately equal to the following concept: ``a tiny change in position, so small that it is well beyond the measurement capabilities of our experimental devices, divided by a similarly tiny change in time." Phrased differently, we may use the equation

\begin{equation}
\label{smallapprox}
    \frac{dx}{dt} =  \lim_{\Delta t \rightarrow 0} \frac{\Delta x}{\Delta t} \approx \frac{\Delta x}{\Delta t}
\end{equation}

if and only if $\Delta x$ and $\Delta t$ are so exceptionally small that the approximation makes no observable effect on the measured variables. For this reason, we can treat $dx$ as if it is the variable $\Delta x$, and similarly $dt$ can be treated as if it is $\Delta t$, as long as we maintain the understanding that this technique is merely an ultra-precise approximation. Thus, we are free to multiply both sides by $dt$ as if it were a variable.

Note that, although we are making an approximation in an intermediate step in this derivation, the end result will be exact. In other words, an infinitesimally close approximation is ultimately an exact solution.

Equation \ref{separated2} includes $dt$ and $dx$, however we want $t$ and $x$. To achieve this goal, let us integrate both sides.

The falling astronaut is at a position $h$ at time $0$ s and at some arbitrary position $x$ at some arbitrary time $t(x)$. We write the arbitrary time as $t(x)$ to indicate explicitly that it is a function dependent on position. Thus, the left-hand side of the equation, the side with time in it, has limits of integration from $t' = 0$ to $t'= t(x)$. Note that the time variables with a prime $'$ are there to indicate which variables are part of the integration process and which variables should remain in the final equation.

The limits of integration on the right-hand side are from the initial position $x' = h$ to some arbitrary position $x' = x$. Again, the prime indicates that the variable is being evaluated, while the unprimed $x$ should remain in the final equation.

With these limits of integration, equation \ref{separated2} becomes:

\begin{equation}
\label{tobeintegrated}
\int_{t'=0}^{t'=t(x)} -\sqrt{2 G M} \, dt' = \int_{x'=h}^{x'=x} \frac{1}{\sqrt{\frac{1}{x'} - \frac{1}{h}}} \, dx'
\end{equation}

This equation is not yet a solution to the original question, as we do not yet have a  straightforward function for $t(x)$, but it is an important stepping stone towards finding the solution. Once we evaluate both of these integrals, we will be left with an equation which only has known constants and two variables, $t$ and $x$. In other words, evaluating the integral will yield the answer.

\subsection{Evaluating the Left-Hand Side}

The integral for the left-hand side of Equation \ref{tobeintegrated} is straightforward:

\begin{equation}
\label{lefthandside}
    \begin{gathered}
        \int_{t'=0}^{t'=t(x)} -\sqrt{2 G M} \, dt'  \\[10pt]
        = -\sqrt{2 G M} \int_{t'=0}^{t'=t(x)}  dt' \\[10pt]
        = -\sqrt{2 G M} \left[ t' \right]_0^{t(x)} \\[10pt]
        = -\sqrt{2 G M} \left( t(x) - 0 \right) \\[10pt]
        = -\sqrt{2 G M} \; t(x) .
    \end{gathered}
\end{equation}

The left-hand side of the equation now has the function $t(x)$, which is what we are solving for.

\subsection{U-Substitution}

The integral for the right-hand side is a multi-step process. Firstly, we rewrite the equation slightly to make it more clear that a u-substitution can be employed. Secondly, we utilize a u-substitution and show that the equation can be rewritten in terms of two functions called a  ``Beta Function" and an ``Incomplete Beta Function." Thirdly, we evaluate the Beta Function. Finally, we use trigonometric identities to evaluate the Incomplete Beta Function.

For the u-substitution, let us first rewrite the right-hand side of Equation \ref{tobeintegrated} to make it more apparent that we can employ a u-substitution in the first place. We multiply the equation by $1 = \frac{\sqrt{x'}}{\sqrt{x'}}$, which yields:

\begin{equation}
\label{almostsub}
\int_{x'=h}^{x'=x} \frac{1}{\sqrt{\frac{1}{x'} - \frac{1}{h}}} \, dx' = \int_{h}^{x} \frac{\sqrt{x'}}{\sqrt{1 - \frac{x'}{h}}} \, dx' .
\end{equation}

The denominator now has the variable and constant clustered together in one term, $\frac{x'}{h}$. We can simplify that term with the following u-substitution:

\begin{equation}
\label{usub}
   u \equiv \frac{x'}{h} \Rightarrow x' = u \, h .
\end{equation}

The first equality in Equation \ref{usub} will be used for the denominator, while the second equality will be used for the numerator.

We must also find a value for $dx'$ in terms of $u$. We will take the second equality in Equation \ref{usub} and derive with respect to $u$ then multiply by $du$, giving us:

\begin{equation}
   x' = u \, h \Rightarrow \frac{dx'}{du} = h \Rightarrow dx'= h \; du.
\end{equation}

We now have substitutive values for the numerator, denominator, and differential element, but we still need substitutive values for the limits of integration.

From our definition of $u$ in Equation \ref{usub}, we see that when $x' = x$, $u = \frac{x}{h}$. We also see that when $x' = h$, $u = \frac{h}{h} = 1$.

We can now substitute all of our values into Equation \ref{almostsub}:

\begin{equation}
\label{endofusub}
\int_{h}^{x} \frac{\sqrt{x'}}{\sqrt{1 - \frac{x'}{h}}} \, dx' = \int_{u=1}^{u=x/h} \frac{\sqrt{uh}}{\sqrt{1-u}} h \, du.
\end{equation}

Although this integral is still difficult to integrate by straightforward methods, we will explore a method for evaluating it in the next section. In other words, Equation \ref{endofusub} has brought us one step closer to finding an analytical solution.

\subsection{The Incomplete Beta Function}

The goal of this section is to rewrite the integral in Equation \ref{endofusub} in a manner that can be evaluated using analytical techniques, without the use of a calculator or numerical approximations. We begin by writing \ref{endofusub} as

\begin{equation}
\label{almostbeta}
\int_{u=1}^{u=x/h} \frac{\sqrt{uh}}{\sqrt{1-u}} h \, du = h^{3/2}\int_{1}^{x/h} {u^{1/2}(1-u)^{-1/2}} \, du .
\end{equation}

This new way of writing the equation may initially appear to have no more utility than the previous way. However, let us compare Equation \ref{almostbeta} to the Incomplete Beta Function, which is defined as:

\begin{equation}
\label{beta}
B(z;a,b) = \int_0^z u^{a-1} (1-u)^{b-1} du,
\end{equation}

where $a$ and $b$ are constants. The Incomplete Beta Function obeys certain mathematical identities which will help us analytically evaluate the integral -- no calculator or numerical approximations will be needed. If we can rewrite Equation \ref{almostbeta} so that the limits of integration begin with $0$ rather than $1$, Equation \ref{almostbeta} will become an Incomplete Beta Function, and we can finally evaluate our integral and find an analytical relationship between $t$ and $x$.

To rewrite our equation so that it resembles an Incomplete Beta Function, we can note that, for any integral,

\begin{equation}
\int_{a}^{b} f(x) \, dx = \int_{0}^{b} f(x) \, dx - \int_{0}^{a} f(x) \, dx .
\end{equation}

Thus, Equation \ref{almostbeta} can be written as:

\begin{equation}
\label{bigboy}
h^{3/2}\int_{1}^{x/h} {u^{1/2}(1-u)^{-1/2}} \, du= h^{3/2} \left( \int_0^{x/h} u^{1/2}(1-u)^{-1/2} du - \int_0^1 u^{1/2}(1-u)^{-1/2} du \right) .
\end{equation}

We now have two Incomplete Beta Functions, which can be written as $B(\frac{x}{h};\frac{3}{2},\frac{1}{2})$ and $B(1;\frac{3}{2},\frac{1}{2})$.

How did we determine that $a = \frac{3}{2}$? Upon comparing Equation \ref{almostbeta} to \ref{beta}, we see that $u^{1/2}$ corresponds to $u^{a-1}$, and thus

\begin{equation}
    u^{1/2} = u^{a-1} \Rightarrow \frac{1}{2} = a -1 \Rightarrow \frac{3}{2} = a.
\end{equation}
        
We used a similar technique for determining that $b = \frac{1}{2}$.

To review, the goal is to evaluate the integral on the right-hand side of Equation \ref{tobeintegrated}, which will provide us with an analytical solution for the position of the astronaut at a given time. Using a u-substitution and the Incomplete Beta Function, we have rewritten the right-hand side as:

\begin{equation}
\label{TwoBetas}
\int_{x'=h}^{x'=x} \frac{1}{\sqrt{\frac{1}{x'} - \frac{1}{h}}} \, dx' = h^{3/2} \left( B(\frac{x}{h};\frac{3}{2},\frac{1}{2}) - B(1;\frac{3}{2},\frac{1}{2}) \right).
\end{equation}

We are now ready to use mathematical identities that the Incomplete Beta Function obeys and take the next step toward finding an analytical equation for $x$ and $t$.

\subsection{Evaluating B(1; 3/2, 1/2)}

When an Incomplete Beta Function is integrated from $0$ to $1$, it becomes a Complete Beta Function, or simply a Beta Function, denoted as $B(a,b)$. The Beta Function obeys this identity:

\begin{equation}
\label{intermsofgamma}
    B(a,b) = \int_0^1 u^{a-1} (1-u)^{b-1} du = \frac{\Gamma(a)\Gamma(b)}{\Gamma(a+b)},
\end{equation}

where the function $\Gamma(w)$ is called the ``Gamma Function" and is defined by

\begin{equation}
    \Gamma(w) = \int_0^\infty y^{w-1} e^{-y} dy.
\end{equation}

Note that the value $y$ does not have any direct physical meaning, but it nonetheless serves a mathematical purpose. It is a ``dummy variable," a variable which appears in an integral but is evaluated, and thus vanishes from the final result.

This identity may at first appear to make the equation more complicated than before. However, the Gamma Function has been studied extensively by mathematicians, and its values are known for many values of $w$. In our particular case, we are interested in $\Gamma(\frac{3}{2})$, $\Gamma(\frac{1}{2})$, and $\Gamma(\frac{3}{2}+\frac{1}{2}) = \Gamma(2).$ We can look up these values from a published table. In our particular case, we have:

\begin{equation}
    \label{threegammas}
    \begin{gathered}
        \Gamma(\frac{3}{2}) = \frac{\sqrt{\pi}}{2}  \\[10pt]
        \Gamma(\frac{1}{2}) = \sqrt{\pi} \\[10pt]
        \Gamma(2) = 1 \\[10pt]
    \end{gathered}
\end{equation}

Thus, our Beta Function simplifies to 

\begin{equation}
\label{piover2}
    B(\frac{3}{2}, \frac{1}{2}) = \frac{\Gamma\left(\frac{3}{2}\right)\Gamma\left(\frac{1}{2}\right)}{\Gamma\left(\frac{3}{2}+\frac{1}{2}\right)} = \frac{\sqrt{\pi}/2 \times \sqrt{\pi}}{1} = \frac{\pi}{2}.
\end{equation}

What a remarkable result! Despite the lengthy derivation involving complex integrals and tortuous substitutions, the end result is exceptionally succinct: $\frac{\pi}{2}$. Let us now evaluate the other Incomplete Beta Function in Equation \ref{TwoBetas}.

\subsection{Evaluating B(x/h; 3/2, 1/2)}

Although this integral will not yield a result as elegant as in the previous section, it will yield a manageable, analytical solution relying only on standard operations such as trigonometric functions and square roots.

To evaluate the $B(\frac{x}{h};\frac{3}{2},\frac{1}{2})$, we will use the following mathematical identity that the Incomplete Beta Function obeys:

\begin{equation}
\label{trigidentity1}
    B(z;a,b) = \int_0^z u^{a-1} (1-u)^{b-1} du = 2 \int_0^{\sin^{-1}\left(\sqrt{z}\right)} \sin^{2a-1}(\theta) \cos^{2b-1}(\theta) d\theta.
\end{equation}

As with the variable $y$ in the previous section, the variable $\theta$ is a dummy variable that appears only in intermediate steps and will disappear from the equation by the time the final result is reached.

This trigonometric identity, along with Equation \ref{intermsofgamma}, is the primary motive behind introducing the Incomplete Beta Function. It will help us integrate the equation and produce a typical, analytical equation that will model the motion of the falling astronaut. Although Equation \ref{trigidentity1} may appear intimidating, we will see that our particular values of $a$ and $b$ cause the equation to simplify into a more manageable form.

Inserting our values for $z$, $a$, and $b$ into Equation \ref{trigidentity1}, our function becomes

\begin{equation}
    B\left(\frac{x}{h}; \frac{3}{2}, \frac{1}{2}\right) = 2 \int_0^{\sin^{-1}\left(\sqrt{x/h}\right)} \sin^2(\theta) \cos^0(\theta) d\theta = \int_0^{\sin^{-1}\left(\sqrt{x/h}\right)} 2\sin^2(\theta) d\theta,
\end{equation}

noting that cos$^0(\theta) =1$ and that we have brought the factor 2 inside the integral.

The antiderivative of sin$^2(\theta)$ is not intuitive, however we can use the following trigonometric identity to make the antiderivative easier to find:

\begin{equation}
    2\sin^2(\theta) = 1 - \cos(2\theta) .
\end{equation}

We make the substitution and evaluate the integral:

\begin{equation}
\label{evaluated}
    \begin{gathered}
        B\left(\frac{x}{h}; \frac{3}{2}, \frac{1}{2}\right) = \int_0^{\sin^{-1}\left(\sqrt{x/h}\right)} \big( 1 - \cos(2\theta) \big) d\theta  \\[10pt]
        = \left[\theta - \frac{\sin(2\theta)}{2}\right]_0^{\sin^{-1}\left(\sqrt{x/h}\right)} \\[10pt]
        = \sin^{-1}\left(\sqrt{x/h}\right) - \frac{1}{2}\sin\left(2\sin^{-1}(\sqrt{x/h)}\right), \\[10pt]
    \end{gathered}
\end{equation}


We did it! We have an analytical solution to $B\left(\frac{x}{h}; \frac{3}{2}, \frac{1}{2}\right)$. However, we can simplify this even further. The second term has an inverse sine function within the sine, which is rather inelegant. However, there exists the following trigonometric identity to remove this inelegancy:

\begin{equation}
    \sin\left(2\sin^{-1}(m)\right) = 2m\sqrt{1-m^2}.
\end{equation}

Substituting this identity into Equation \ref{evaluated}, and noting that $m = \sqrt{x/h}$ in this case,

\begin{equation}
\label{secondIBF}
    B\left(\frac{x}{h}; \frac{3}{2}, \frac{1}{2}\right) = \sin^{-1}\left(\sqrt{x/h}\right) - \sqrt{x/h} \sqrt{1 - x/h}.
\end{equation}


We have finally evaluated both Incomplete Beta Functions. We will use them to evaluate the right-hand side of equation \ref{tobeintegrated} in the next section and, at long last, determine our equation of motion for the falling astronaut.

\subsection{Bringing Everything Together}

To review, our goal is to model the motion of an astronaut as she falls from a height $h$ to a new position $x$ over time, noting that the acceleration due to gravity increases as the astronaut approaches the planet and thus standard kinematic equations are not applicable.

We used conservation of energy and the ``Separation of Variables" technique to derive Equation \ref{tobeintegrated}, which has an integral on both sides. We evaluated the left-hand side of Equation \ref{tobeintegrated} with standard techniques, finding it to be $-\sqrt{2 G M} \; t(x)$ in Equation \ref{lefthandside}.

For the right-hand side, we rewrote the integral in terms of two Incomplete Beta Functions in Equation \ref{TwoBetas}. We calculated that the second Incomplete Beta Function is simply $\frac{\pi}{2}$ in Equation \ref{piover2}, and we found that the first Incomplete Beta Function can be written in terms of an inverse sine function and square roots in Equation \ref{secondIBF}.

Now, substituting $-\sqrt{2 G M} \; t(x)$ into the left-hand side of Equation \ref{tobeintegrated}, and substituting our values for the Incomplete Beta Functions into the right-hand side of the equation, the equation becomes:

\begin{equation}
\label{penultimate}
    \begin{gathered}
    -\sqrt{2GM} \; t(x) = h^{3/2}\left(B(\frac{x}{h}; \frac{3}{2}, \frac{1}{2}) - B(\frac{3}{2}, \frac{1}{2})\right)
    \\[10pt]
    \sqrt{2GM}\; t(x) = -h^{3/2}\left( \sin^{-1}\left(\sqrt{x/h}\right) - \sqrt{x/h} \sqrt{1-x/h} - \frac{\pi}{2} \right).
    \end{gathered}
\end{equation}

Finally, we divide by $\sqrt{2GM}$ and distribute the $\sqrt{x/h}$ in the second term on the right-hand side. This yields our function for the time at a given position for the falling astronaut:

\begin{equation}
\label{finally!}
    \boxed{
    t(x) = \frac{h^{3/2}}{\sqrt{2GM}} \left( \frac{\pi}{2} + \sqrt{\frac{x}{h}-\frac{x^2}{h^2}} - \sin^{-1}\left(\sqrt{\frac{x}{h}}\right) \right).
    }
\end{equation}

Our time, dedication, and grit has paid off! For any given position that the astronaut finds herself in, we can calculate how much time has passed using only standard mathematical operations without approximation.

\subsection{The Answer}

The original two questions were:

\textit{(a)} How long does it take for the astronaut to fall to the surface of the planet from her initial height?

\textit{(b)} What is the astronaut's elevation $x$ for any given time $t$?

To calculate how much time passes between being released and arriving at the surface of the Earth, we can use Equation \ref{finally!} to compute the final time minus the initial time, or

\begin{equation}
\Delta t = t_f - t_i = t(R) - t(h).
\end{equation}

The problem statement tells us that $t_i$ should be $0$. Let us check that the time at the moment of being released, $t(h)$, truly yields 0:

\begin{equation}
\label{timezero}
    \begin{gathered}
    t(h) = \frac{h^{3/2}}{\sqrt{2GM}} \left( \frac{\pi}{2} + \sqrt{\frac{h}{h}-\frac{h^2}{h^2}} - \sin^{-1}\left(\sqrt{\frac{h}{h}}\right) \right)
    \\[10pt]
    = \frac{h^{3/2}}{\sqrt{2GM}} \left( \frac{\pi}{2} + \sqrt{1-1} - \sin^{-1}\left(1\right) \right) = \frac{h^{3/2}}{\sqrt{2GM}} \left( \frac{\pi}{2} + 0 - \frac{\pi}{2} \right) = 0.
    \end{gathered}
\end{equation}

The equation does indeed yield zero! Thus, the time required to fall from the initial height to the final height is simply $t(R)$.

Using the numerical values provided in the  problem statement, we can input Equation \ref{finally!} into a calculator and determine $t_f$. The calculator provides the values

\begin{equation}
\label{answer}
    \boxed{t_f=1263 s}
\end{equation}

with four significant figures. In other words, the astronaut would fall for slightly more than twenty-one minutes. We can suppose that the astronaut lands safely onto a very soft material.

With regard to part $(b)$, notice that Equation \ref{finally!} give us a function for time in terms of position, but not position in terms of time. Is it possible to rewrite the equation so that $x$ is isolated on one side of the equation, and the other side is written only in terms of constants and the variable $t$?

The short answer is: no, not by conventional methods. The variable $x$ appears both inside and outside of an inverse sine function, which makes isolating $x$ exceptionally difficult, even by the standards of the derivation that just occurred. There is simply no simple way to write $x(t)$ as a compact, closed-form function in terms of $t$.

\begin{figure}[ht]
\centering
\includegraphics[width=0.5\textwidth]{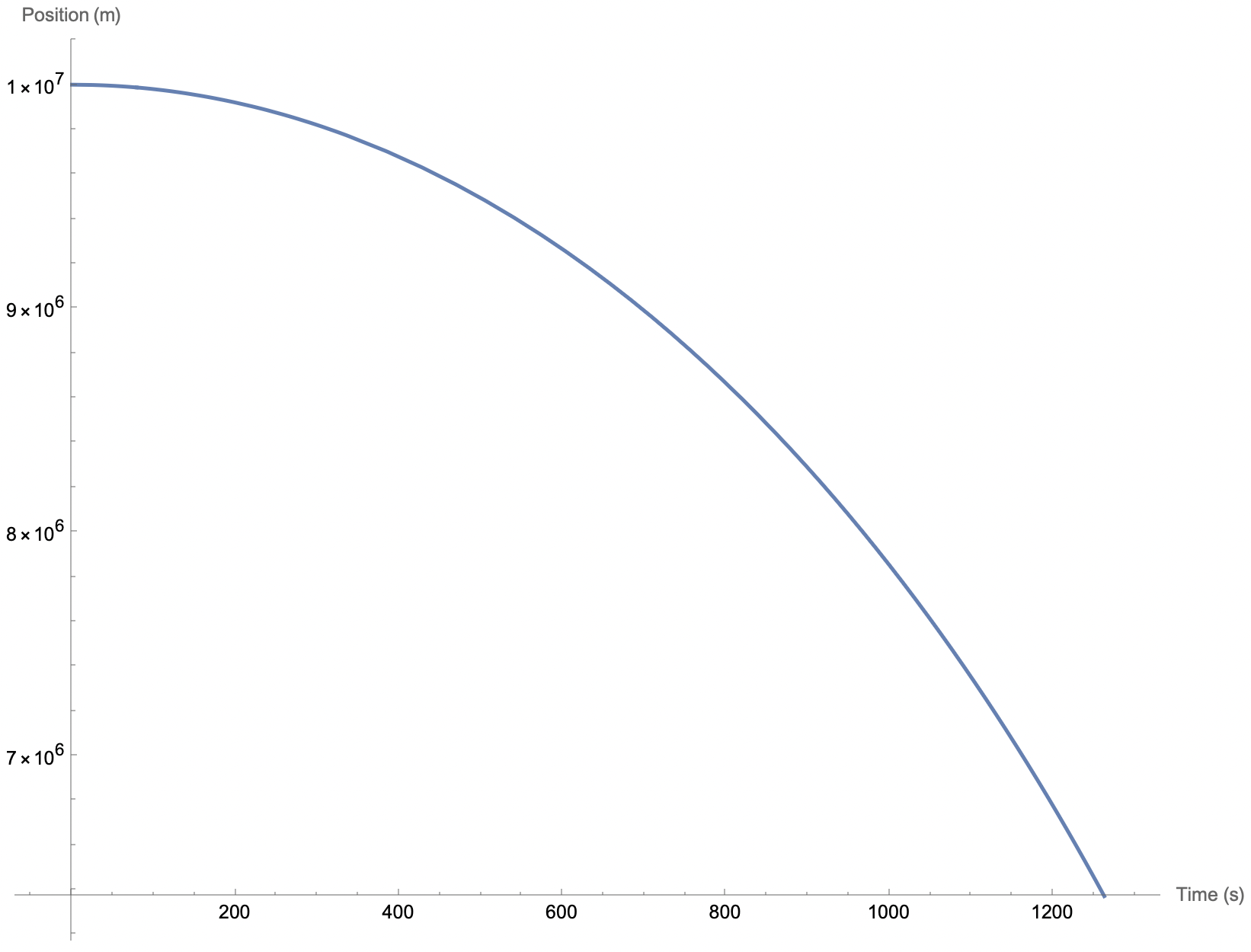}
\caption{\label{fig:D1} The time elapsed since the beginning of the astronaut's fall as given by Equation \ref{finally!} is plotted on the horizontal axis as a function of position plotted on the vertical axis. Note that the function is not a parabola, and a quadratic function does not adequately approximate the function.}
\label{graph}
\end{figure}

However, we are free to graph $t(x)$ with time on the horizontal axis and position on the vertical axis, as shown in Figure \ref{graph}.

This graph  provides an answer to part $(b)$. For any time $t$, we may inspect the graph and observe at which position the astronaut is located at that time.

\subsection{Implications}

This derivation is inappropriately advanced for a first-semester undergraduate student. It requires the use of equations which first-semester undergraduates may not be aware of, namely the Incomplete Beta Function, its relationship to the Gamma function, and sophisticated trigonometric identities. Similarly, it relies on the Separation of Variables technique which is not always included in high-school calculus curriculum.

The derivation also partially relies on hindsight. For example, it is not obvious that the right-hand side of \ref{tobeintegrated} would benefit from a u-substitution, nor is it obvious that one should multiply the fraction by $1 = \frac{\sqrt{x}}{\sqrt{x}}$ to reveal the possibility of such a substitution. For these reasons, this analytical solution should be reserved for upper-division classical mechanics courses.

The analytical solution is also incomplete, as we arrived at the function $t(x)$ but not $x(t)$. If a physicist wishes to know the location of the astronaut at a given time, he must approximate the position by inspecting the graph. This approximation defeats the original motive for finding an exact analytical solution.

In the next section, we present a separate solution to the Falling Astronaut Problem, one which relies on differential equations, numerical approximations, and computational methods using a digital spreadsheet. Although this technique may seem advanced for a first-semester undergraduate, the central claim of this paper is that familiarity with high-school level calculus is the only prerequisite for mastery over the technique.

\section{Numerical Approximation}
\label{NumericalApproximation}

\subsection{Setting up our Differential Equation}

The following technique appeals to two equations: Newton's Second Law of Motion and Newton's Law of Universal Gravitation. Newton's Second Law of Motion states that the force $F$ applied to the object is equal to the mass $m$ of the object multiplied by the object's acceleration $a$:

\begin{equation}
\label{NewtonSecondLaw}
    F = m a = m \frac{d^2x}{dt^2}\; .
\end{equation}

Note that the mathematical definition of the acceleration of an object is the second derivative of the object's position with respect to time.

Newton's Universal Law of Gravity states that any two objects of masses $M$ and $m$, respectively, exert a gravitational force on each other that depends on the universal gravitational constant $G$ and the distance between the objects $x$ in the following manner:

\begin{equation}
\label{LawOfGravity}
    F = -\frac{ G M m}{x^2} \; .
\end{equation}

Note that the negative sign indicates that the objects are pulled toward each other.

For the Falling Astronaut Problem, the force causing the astronaut to accelerate downward is the gravitational force. Thus, we can set equations \ref{NewtonSecondLaw} and \ref{LawOfGravity} equal to each other:

\begin{equation}
    \label{needapproximation}
    \begin{gathered}
        m \frac{d^2x}{dt^2} = -\frac{ G M m}{x^2}   \\[10pt]
        \Rightarrow \frac{d^2x}{dt^2} = -\frac{ G M}{x^2}   \\[10pt]
    \end{gathered}
\end{equation}

As with the analytical solution, we see that the mass of the astronaut is irrelevant.

As a brief aside, consider this question: what is the physical meaning behind Equation \ref{needapproximation}? This equation provides us with the acceleration due to gravity -- in other words, $\frac{d^2x}{dt^2}$ = $g$. The acceleration is not constant, but depends on the distance $x$ between the Earth and the astronaut. When we substitute $G$, $M$, and $x = R$ for the numerical values provided in the problem statement, we calculate the value $-9.8$ m s$^{-2}$, the standard value given by high-school physics textbooks for the acceleration due to gravity on the surface of the Earth.

Let us return to our main goal of describing the position over time of the falling astronaut. Equation \ref{needapproximation} is difficult to evaluate -- it is another differential equation! But this equation must be evaluated using a separate technique compared to the previous section. The differential equation we saw in Equation \ref{separated2} was a ``separable" differential equation -- we could rewrite it so that each variable was isolated on one side of the equation with constants. In our case, time appeared only on the left side and position appeared only on the right.

Equation \ref{needapproximation}, in contrast, is ``inseparable" -- there is no straightforward way to isolate $t$ and $x$ on separate sides of the equation. We might be tempted to use a similar technique we used in deriving Equation \ref{separated2} for this new equation, namely multiplying both sides by $dt^2$ and $x^2$. However, we are not mathematically allowed to multiply by $dt^2$ the way we multiplied by $dt$ in Equation \ref{separated2}. With Equation \ref{separated2}, we multiplied $\frac{dx}{dt}$ by $dt$ and were left with $dx$. The term ``$dx$," we argued, has a concrete physical meaning: it is approximately equal to a small change in position, $\Delta x$, in the limit that $\Delta t$ approaches zero. However, when we multiply $\frac{d^2x}{dt^2}$ by $dt^2$, we are left with $d^2x$, or $ddx$. Such a term does not have any direct physical meaning, and it cannot be approximated as being equivalent to some observable quantity, even in the limit of $\Delta t$ approaching zero. For that reason, we cannot separate $d^2x$ from $dt^2$.
 
Despite the differential equation having inseparable variables, there still exists a technique that allows us to solve both parts of the original Falling Astronaut Problem. To find a solution, we must employ a ``numerical approximation."

\subsection{What is a Numerical Approximation?}

Before we answer the question posed in the title of this subsection, let us consider the following conundrum. The main difficulty behind the Falling Astronaut Problem is that the acceleration due to gravity, $g$, is not constant. However, high school physics textbooks often cite that $g$ is in fact a constant $-$9.8 m s$^{-2}$. Why is this?

The resolution to this apparent contradiction is that $g$ changes its value very slowly over vertical distances. Indeed, $g$ is slightly lower in magnitude at the top of a skyscraper than at sea level, however the distance is small enough that it can be ignored. When a physics textbook states that $g$ is constant, the author means that $g$ is approximately constant for any experiments that a student is likely to conduct or see in the classroom.

\begin{figure}[h]
    \centering
    \includegraphics[width=0.9\linewidth]{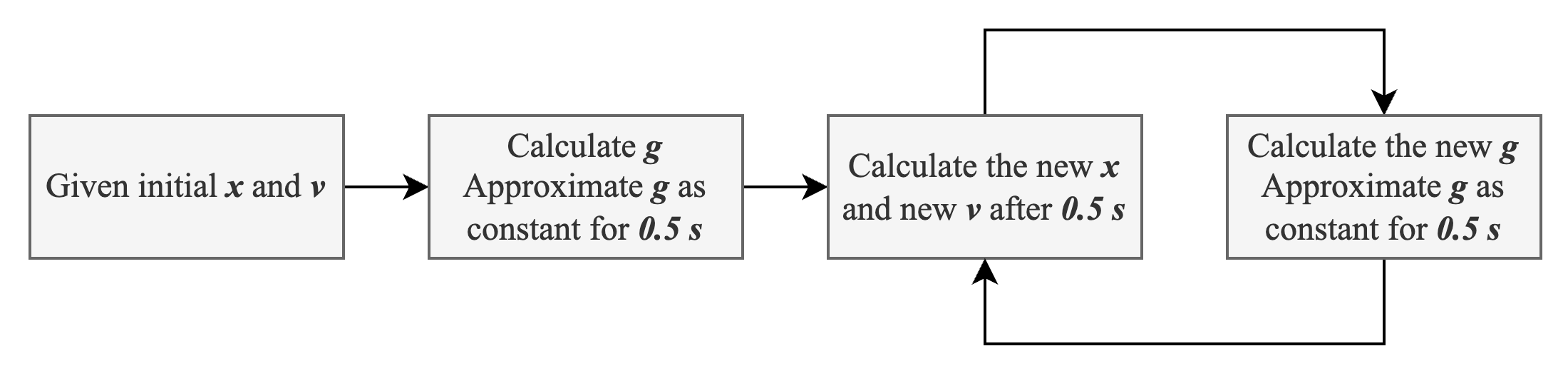}
    \caption{The state of the astronaut at each moment in time dictates the state of the astronaut at a moment in time 0.5 seconds later, which in turn dictates the state of the astronaut 0.5 seconds after that, and so on until the astronaut reaches the surface of the planet.}
    \label{NumApprox}
\end{figure}

For the Falling Astronaut Problem, can approximate $g$ to be constant over both short spatial intervals and short time intervals. Indeed, $g$ will not change substantially over a brief duration of time such as 0.5 seconds. This is the key insight behind the numerical approximation technique we will use in this paper. For any given position of the astronaut, we can calculate $g$ at that position using Equation \ref{needapproximation}. We then assume that $g$ is approximately constant for the next 0.5 seconds. With this known and approximately constant acceleration, we can calculate how far the astronaut moves over the course of 0.5 seconds. In other words, we calculate her new position at the end of that time interval. At that point, we use the new value of position to calculate the new acceleration, which we approximate as constant for 0.5 more seconds, and the cycle repeats. A diagram visualizing the process is shown in Figure \ref{NumApprox}.

In general, a ``numerical approximation" is any mathematical technique that relies on specific numerical values to approximate the solution in lieu of finding general analytical equations with variables, as we did in the previous section. The particular numerical approximation described in this section has features in common with ``Euler's Method," a technique named after the mathematician Leonhard Euler, who described the method in the eighteenth century.

\subsection{Setting up our Numerical Approximation}
\label{SettingUp}

We are not yet ready to implement our numerical approximation for Equation \ref{needapproximation}. The equation as written is ``continuous," that is, it applies to any value of $x$ and its corresponding $t$. In contrast, we are only interested in specific and individual values of $x$, $v$, and $g$ at particular moments in time, namely once every 0.5 seconds. We will need to rewrite Equation \ref{needapproximation} so that it takes on values that are ``discrete," that is, individually separate and distinct.\footnote{To provide an analogy, picture two different car radios. For the first car radio, the volume knob turns smoothly, and you can set the volume to any value. In the second car, the volume knob clicks when you turn it, and it can only take on specific volume levels as indicated by whole numbers one through ten around the dial. The volume of the first radio is continuous, while the volume of the second knob is discrete.}

We will need to ``discretize" Equation \ref{needapproximation} by transforming it from a continuous function to a discrete function. To do this, we must discretize the definitions of the first and second derivatives.

The mathematical definition of a derivative of a function $f$ with respect to a variable $z$ is:
\begin{equation}
\label{derivative}
    \frac{d}{dz}f(z) \equiv \lim_{h \to 0} \frac{f(z+h) - f(z)}{(z+h) - z} = \lim_{h \to 0} \frac{f(z+h) - f(z)}{h},
\end{equation}

and the mathematical definition of velocity is:

\begin{equation}
    v(t) = \frac{d}{dt}x(t) ,
\end{equation}

thus we can write the velocity at a given moment in time as:

\begin{equation}
\label{inelegant}
    v(t) = \lim_{\Delta t \to 0} \frac{x(t+\Delta t) - x(t)}{(t+\Delta t) - t} = \lim_{\Delta t \to 0} \frac{x(t+\Delta t) - x(t)}{\Delta t} .
\end{equation}

The current manner in which we are writing our variables will become cumbersome in the following derivation. To make our derivation more succinct, we will rename the following variables. Note that we are not introducing any new concepts -- we are simply renaming our terms:

\begin{equation}
    \begin{gathered}
        v(t) \equiv v_{n} \\[10pt]
        x(t) \equiv x_n \\[10pt]
        x(t+\Delta t) \equiv x_{n+1} \; .
    \end{gathered}
\end{equation}

What is the physical meaning of $n$? Our values of position and velocity will soon be discretized, so there exists some first value for position and velocity, some second value, a third value, and so on. We could list each value for $x$ and $v$ and give each entry on the list number, starting with 1, then 2, and so on. The subscript $n$ indicates the entry number for $x$ and $v$ on that list. For example, $x_1$ and $v_1$ are the first entries on our list of discrete values, while $x_{22}$ and $v_{22}$ appear twenty-second on the list. The subscript $n+1$ indicates the particular value immediately after some other given value.

Equation \ref{inelegant} can now be written in a more succinct manner:

\begin{equation}
    v_n = \lim_{\Delta t \to 0} \frac{x_{n+1}- x_n}{\Delta t} .    
\end{equation}

Now for the approximation. Equation \ref{answer} found that the astronaut's fall lasts for approximately 1263 seconds. Thus, a time duration such as 0.5 seconds is very short compared to the duration of the whole journey. If we maintain a very small value of $\Delta t$ = 0.5 s, we can make the same approximation as we did in Equation \ref{smallapprox}:

\begin{equation}
\label{v_n}
    v_n = \lim_{\Delta t \to 0} \frac{x_{n+1}- x_n}{\Delta t} \approx \frac{x_{n+1} - x_{n}}{\Delta t} .
\end{equation}

Conceptually, this equation tells us that the astronaut's velocity is approximately constant over a short duration of space and time.

We can make a similar derivation to write an equation for $v_{n-1}$, the discrete velocity of the astronaut that occurs one step in time prior to $v_n$:

\begin{equation}
\label{v_n-1}
    v(t-\Delta t) \equiv v_{n-1} \approx \frac{x_{n} - x_{n-1}}{\Delta t} .
\end{equation}

Notice that our approximation has discretized the velocity, position, and time. Time now occurs in steps of 0.5 seconds. In this approximation, time flows like that of an old-fashioned film with a low frame rate, where motion occurs in specific jumps rather than flowing smoothly.

Similarly, we can no longer plug in any arbitrary value for $x$ or $v$. As previously stated, we must instead imagine the values of position and velocity as being numbered entries on a list, where $n$ is their entry number.

We can also discretize the acceleration, which is the time derivative of velocity:

\begin{equation}
\label{acceleration}
    a(t) = \frac{d}{dt}v(t).
\end{equation}

Note that the acceleration is both time derivative of velocity and the second time derivative of position -- the two definitions are equivalent.

We can also define the derivative of a function in a way that is slightly different from Equation \ref{derivative} but is equally valid mathematically:

\begin{equation}
\label{newderivative}
    \frac{d}{dz}f(z) \equiv \lim_{h \to 0} \frac{f(x) - f(z-h)}{z - (z-h)} = \lim_{h \to 0} \frac{f(z) - f(z-h)}{h}.
\end{equation}

In the context of acceleration as the time derivative of velocity, we can combine equation \ref{acceleration} and \ref{newderivative} to acquire:

\begin{equation}
    a(t) = \lim_{\Delta t \to 0} \frac{v(t) - v(t-\Delta t)}{\Delta t} = \lim_{\Delta t \to 0}\frac{v_{n} - v_{n-1}}{\Delta t}.
\end{equation}

For small \(\Delta t\), we can make the same approximation as before:

\begin{equation}
\label{approximatea}
    a(t) \approx \frac{v_{n} - v_{n-1}}{\Delta t} .
\end{equation}

Next, we can take our values for $v_n$ and $v_{n-1}$ given in Equation \ref{v_n} and Equation \ref{v_n-1} and substitute them into Equation \ref{approximatea}:

\begin{equation}
\label{discreteacceleration}
    a \approx (v_{n} - v_{n-1})\frac{1}{\Delta t}=\left( \frac{x_{n+1} - x_n}{\Delta t} - \frac{x_n - x_{n-1}}{\Delta t} \right) \frac{1}{\Delta t} = \frac{x_{n+1} - 2x_n + x_{n-1}}{\Delta t^2}.
\end{equation}

We have discretized the acceleration! Let us return to Equation \ref{needapproximation}, the differential equation. We can discretize it by replacing $x$ with $x_n$ and replacing the acceleration ($a=\frac{d^2x}{dt^2}$) with the discretized value in equation \ref{discreteacceleration}:

\begin{equation}
\label{discretizedgravity}
   \frac{x_{n+1} - 2x_n + x_{n-1}}{\Delta t^2} \approx  -\frac{G M}{x_n^2}
\end{equation}

The core goal of this numerical approximation is to determine the next discrete position in our list of entries. In other words, if we know the current position $x_n$ and the previous position $x_{n-1}$, our goal is to determine the next position $x_{n+1}$. Thus, let us rearrange our equation so that $x_{n+1}$ is isolated. Algebraically rearranging Equation \ref{discretizedgravity}, we acquire

\begin{equation}
\label{key}
    \boxed{
    x_{n+1} \approx 2x_n - x_{n-1} - \frac{G M \Delta t^2}{x_n^2} .}
\end{equation}

Equation \ref{key} is the key equation of this section. For every position $x_n$ and previous position $x_{n-1}$, we can determine the next position $x_{n+1}$ of the astronaut. We use Equation \ref{key} to solve the Falling Astronaut Problem in the next section.

\subsection{Computing the Approximation}
\label{Computing}

We can only utilize our numerical approximation if we know the initial conditions of the astronaut. In our case, we are told that the initial position is $1.0000 \times 10^7$ m and that the initial velocity is 0 m s$^{-1}$. Thus, we can comfortably say that the initial position, $x_1$, is

\begin{equation}
    x_1=1.0000 \times 10^7 \; \text{m} \;.
\end{equation}

What is $x_2$? Upon inspecting Equation \ref{key}, we see that if $x_n = x_1$, then $x_{n+1} = x_2$. However, there is a third term: $x_{n-1}$, which based on our previous reasoning must be $x_0$.

To find the value of $x_0$, we must use our other initial condition: that the initial velocity, which we will name $v_0$, is zero. Using Equation \ref{v_n-1}, we see that:

\begin{equation}
    v_0 = 0 \; \text{m s}^{-1} = \frac{x_1 - x_0}{\Delta t} = \frac{1.0000 \times 10^7 - x_0}{\Delta t} \Rightarrow x_0 = 1.0000 \times 10^7 \; \text{m} \;.
\end{equation}

Conceptually, this answer makes sense: the velocity is zero for the first 0.5 seconds, so after 0.5 seconds, the position has not changed, and the value for $x_1$ is identical to $x_0$.

Substituting our values for $G$, $M$, $x_{1}$, and $x_{0}$ into Equation \ref{key}, we calculate:

\begin{equation}
    x_{2}= 9,999,999 \; \text{m} \;.
\end{equation}

After a subsequent 0.5 seconds (for a total of $t = 1.0$s), the astronaut falls a mere one meter.

We can repeat this process to find $x_3$. Originally, $x_2 = x_{n+1}$, $x_1=x_n$, and $x_0=x_{n-1}$. However, we are no longer interested in calculating the position at $t$ = 1.0 s based on the positions at $t$ = 0.5 s and $t$ = 0 s. Instead, we are calculating the position at $t$ = 1.5 s based on the positions at $t$ = 1.0 s and $t$ = 0.5 s. Thus, in our new calculation, $x_3 = x_{n+1}$, $x_n=x_2$, and $x_{n-1}=x_1$. We plug in these new values into Equation \ref{key} to yield:

\begin{equation}
    x_3 = 9,999,997 \; \text{m} \;.
\end{equation}

After 1.5 seconds, the astronaut has fallen three meters.

We can continue this process as many times as we need, each position dictating the next, until the astronaut reaches the surface of the planet. We can use computer software such Google Sheets, a digital spreadsheet, to repeat this process thousands of times. Using this software, we eventually reach:

\begin{equation}
    \begin{gathered}
    \label{Iterated}
        x_{2526} = 6,373,564 \; \text{m} \quad \\[10pt]
        x_{2527} = 6,370,195 \; \text{m} \; \; .
    \end{gathered}
\end{equation}

Notice that $x_{2527}$ is less than the radius of the planet $R$, while $x_{2526}$ is greater than $R$. The astronaut must have landed at some time between the two moments in time corresponding to $x_{2556}$ and $x_{2527}$, respectively.

How can we calculated the time $t$ for a given position $x_n$? Let us consider the following: the first iteration of Equation \ref{key} corresponds to 0 seconds, and each successive iteration of Equation \ref{key} corresponds to 0.5 seconds passing. Thus, the amount of time that passes after 2526 iterations is:

\begin{equation}
    0.5 \; \frac{s}{\text{iteration}} \times ( 2526 \; \text{iterations} -1 \; \text{iteration})= 1262.5 \;s \;.
\end{equation}

Note that we must subtract 1 iteration from the number of iterations so that the first iteration corresponds to 0 seconds, not 0.5 seconds.

Similarly, the amount of time that passes of 2527 iterations is

\begin{equation}
    0.5 \; \frac{s}{\text{iterations}} \times 2527 \; \text{iterations} = 1263.0 \; s \; .
\end{equation}

Thus, the amount of time that passes between the astronaut being released and the astronaut reaching the surface is somewhere between $t$ = 1262.5 s and $t$ = 1263.0 s. Based on the numerical values provided in the problem statement, our final answer ought to have four significant figures. Thus, our answer is:

\begin{equation}
    \boxed{t_f=1263 s .}
\end{equation}

This answer is identical to the analytical solution! These two seemingly unrelated techniques yield the same answer. 

\begin{figure}[ht]
    \centering
    \includegraphics[width=0.75\linewidth]{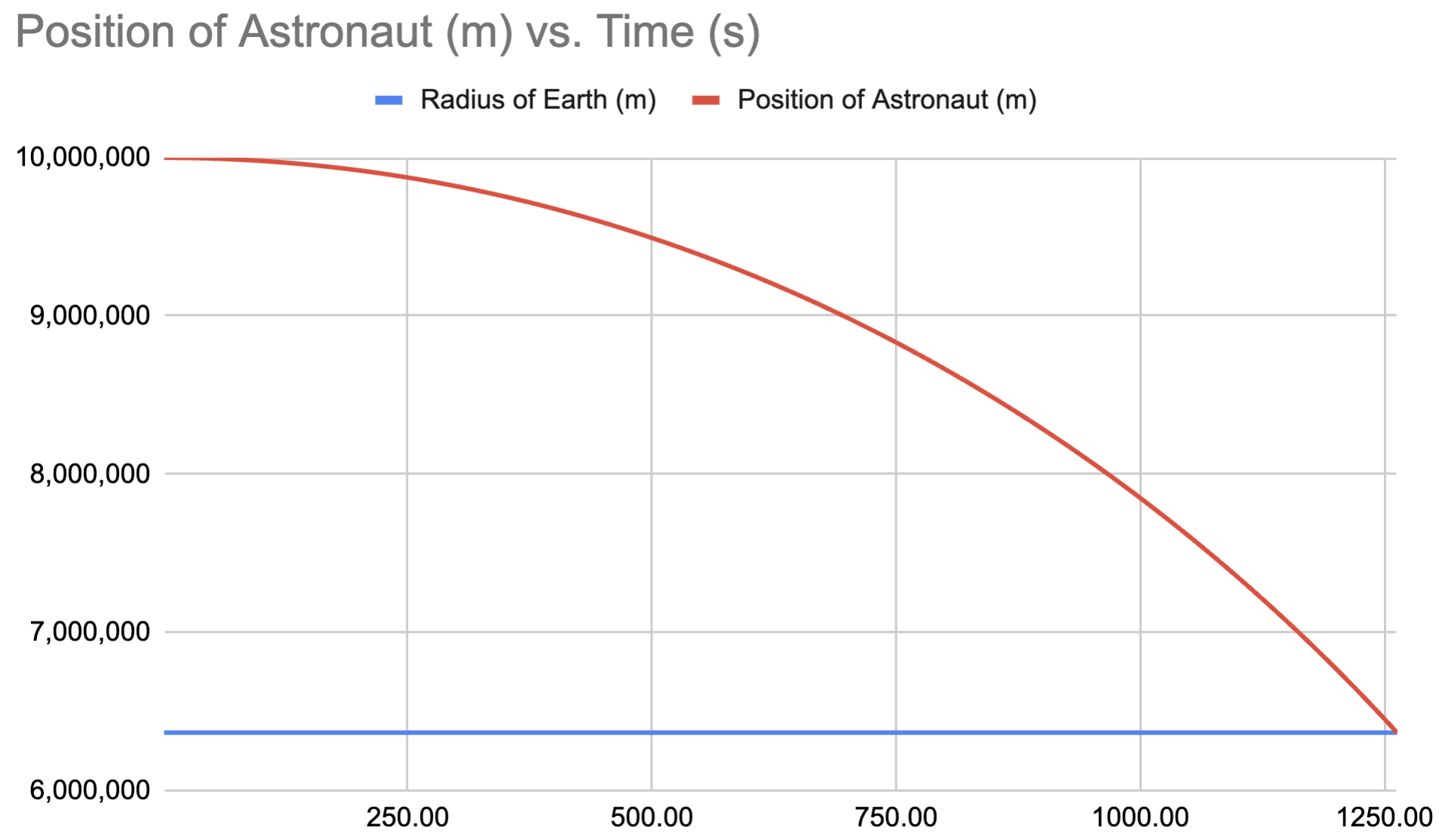}
    \caption{The position of the astronaut above the center of the Earth is plotted against time using the Euler method, a numerical approximation.}
    \label{numericalgraph.png}
\end{figure}

We can also plot every position value $x_n$ with its corresponding time $t$ on the horizontal axis, as shown in Figure \ref{numericalgraph.png}. As in the previous section, this graph constitutes an answer to part $(b)$ of the question. However, we do not need to read the data directly from the graph. Rather, for any time $t$, we can iterate Equation \ref{key} an appropriate number of times until we arrive at the corresponding $x_n$. In this way, we can find the astronaut's position for any given time.

Note that Figure \ref{numericalgraph.png} provides approximately the same data as in Figure \ref{graph}. As in part $(a)$, both the analytical derivation and the numerical approximation yield approximately the same results.

\subsection{Implications}

Numerical approximations are complex, but the core assertion in this paper is that the numerical approximations are within the reach of first-semester undergraduate students. Indeed, the particular numerical approximation described in this section only requires familiarity with Newton's Second Law of motion, Newton's Law of Universal Gravitation, acceleration and velocity as the second and first derivatives of position  (respectively), a willingness to learn the distinction between discrete and continuous equations, and the determination necessary for learning how to use spreadsheets.

As a point of comparison, the analytical derivation requires an understanding of integrals, u-substitutions, a relatively obscure mathematical function, and sophisticated trigonometric identities. The numerical approximation uses none of these.

The main barrier towards utilizing numerical approximations is most likely not in the complexity of the mathematics, but in the difficulty of conceptualizing discretized time and the iterative nature of equations such as Equation \ref{key}. This particular numerical approximation lies behind a conceptual barrier, not a mathematical one, and thus this paper urges teachers to introduce their students to this branch of mathematics and help students garner an appreciation for the power of numerical evaluations of differential equations.

\section{Implementing Numerical Approximations in the Classroom}
\label{Implementation}

This section outlines a classroom laboratory exercise intended for first-year undergraduate physics students with high-school level experience in forces, gravity, and kinematics. It is intended to take 1-3 hours, assuming that the students will need time to familiarize themselves with the software, mentally adjust to thinking about time and position as a discrete quantity, and ask questions about the process. For example, they may wonder why we are unable to use the familiar kinematic equation stated in Equation \ref{Kinematic}.

The laboratory exercise below provides step-by-step instructions for using Google Sheets to iterate Equation \ref{key} thousands of times to yield Equation \ref{Iterated}. We also provide a potential homework assignment which utilizes the same technique in a new context. The goal of the homework assignment is to help the students conceptualize the method presented in this section rather than mechanically copying the instructions. 

This exercise is intended to be done after students have had one class on the Falling Astronaut Problem, in which they are exposed to the problem statement as described in Section \ref{ProblemStatement} and have been given a class-long overview of the technique described in Section \ref{NumericalApproximation}. However, these instructions assume that students are new to the concept of a numerical approximation and will likely engage in productive struggle as they familiarize themselves with this technique. Indeed, the activity is intended to be challenging, but not so challenging as to be beyond the reach of the undergraduate students.

\subsection{Getting Started and Defining the Time Column}
Open Google Sheets online and click ``Blank Spreadsheet." The vertical columns are labeled with letters, with A starting on the far left proceeding to B and onward to the right. The horizontal rows are numbered, with 1 at the top and proceeding to 2 and onward downward.

The goal is to fill the first two columns with data by iterating Equation \ref{key}. Let us label these two columns. Click on box A1 and type ``Time," then click on box B1 and type ``Position."

In Box A2, write ``0," as our initial time is assumed to be $t = 0$ s. In the box below, write:

\begin{equation}
    \label{timestep}
    =A2+0.5 \quad.
\end{equation}

Press ``Enter" on Windows or ``Return" on a Mac when the equation has been written out.

Note that the period is not part of the equation.

Equation \ref{timestep} tells the computer to take the value in A2 and add 0.5. In our particular case, the box returns a value that is 0.5 greater than the box immediately above it. Note that the equality sign tells the computer that the entry is a mathematical value, not plain text.

Upon clicking on A3, there will be a blue highlight around the box and a small dot in the bottom-right corner. Clicking on the box and dragging the mouse directly downward will cause  the boxes included within the dragging motion to acquire the a similar equation, that of stating a value 0.5 greater than the value in the box immediately above.

Suppose we want thousands of boxes in Column A to obey this equation. We can scroll to the bottom of the spreadsheet until we see ``Add 1000 more rows at the bottom." Click ``Add" twice. This will give us more than enough rows for the Falling Astronaut Problem, which requires 2527 iterations. Next, click box A3, scroll down to row 3000, and click box A3000 while holding down the ``shift" key. This will highlight the thousands of empty boxes in Column A. Click on the ``Fill Down" option from the toolbar, or alternatively press ``Ctrl + D" on Windows or ``Command + D" on a Mac. This will fill all selected columns with the same equation as A3. Each new row should have a number 0.5 more than the row above it.

\subsection{Defining the Position Column}
Now we fill in the data for position. Write ``10000000" for Boxes B2 and B3, as that is the initial position provided by the problem statement. Using the same reasoning provided in Section \ref{Computing}, the initial velocity is 0 m s$^{-1}$, and thus the position immediately after the initial position is 10,000,000 m as well. Note that we may also write ``=10$^{\wedge}$7", which yields the same value.

The next step is the most difficult part. We wish to write Equation \ref{key} in Box B4 in a format that the Google Sheets software can properly understand We can write the equation as:

\begin{equation}
\label{GoogleSheet}
  =2*\text{B}3-\text{B}2-6.6743*10^\wedge(-11)*5.97219*10^\wedge(24)*0.5^\wedge2 \; / \; \text{B}3^\wedge2  \quad .
\end{equation}

Let us compare Equation \ref{key} with Equation \ref{GoogleSheet}. Although there is nothing written on the left-hand side of the equality in Equation \ref{GoogleSheet}, We are typing this equation into box B4, and thus there is an implied ``B4" on the left-hand side. Box B4 represents $x_{n+1}$, the next discrete position in our iterative process. The box above, B3, represents $x_n$, the most recently calculated position value. Similarly, B2 represent ``$x_{n-1}$," the value immediately prior to $x_n$. Note that in in Equation \ref{GoogleSheet} we have substituted $G$, $M$, and $\Delta t$ for their numerical values as given in the problem statement.

Upon pressing the ``return" or ``Enter" key, Box B4 will compute Equation \ref{GoogleSheet} and return the value of 9999999.003. As with Column A, we can click on Box B4, click on the small blue box in the bottom-right corner, and drag the mouse downward to apply a similar equation to the boxes below B4. Note that the equation defined in each new box only refers to the two boxes immediately above it. Box B20, for example, will be defined by the equation:

\begin{equation}
    =2*\text{B}19-\text{B}18-6.6743*10^\wedge(-11)*5.97219*10^\wedge(24)*0.5^\wedge2 \; / \; \text{B}19^\wedge2  \quad .
\end{equation}

As with Column A, we can fill in the rest of the thousands of boxes in column B by clicking on B4, scrolling down to the bottom of the sheet, clicking on B3000 while holding down the ``shift" key, and clicking on the ``Fill Down" (or ``Command + D" on a Mac and ``Ctrl + D" on a Windows). This will fill in the whole    ``Position" column with data. We now have two columns, one representing the time elapsed since the astronaut began her fall and one representing the position of the astronaut at that time.

\subsection{Trimming our Data}

Let us assume that the astronaut safely comes to a stop upon reaching the surface of the planet, when $x =$ 6371000 m. The astronaut does not continue sinking into the Earth, and thus we can delete the rows which have position values less than 6371000 m. We can scroll down through the values of position until we find a box with a value less than 6371000 m. The first box with a position value less than 6371000 m is box B2528, with a value of 6370195.059. Note that Box B2528 corresponds to the 2527th iteration, not the 2528th iteration, as Row 1 contains the column labels ``Time" and ``Positions" rather than data, making Row 2 the first iteration.

Although Box 2528 does not itself have physical meaning, as it is below 6371000 m, we should keep it in our spreadsheet because the astronaut lands somewhere between iteration 2526 and iteration 2527. Thus, boxes B2527 and B2528 together define a specific window of time in which the astronaut lands.

In contrast, rows 2529 and beyond can be deleted. To delete these rows, click on the number 2529 on the far left side of the screen, scroll down to the number 3000, shift-click ``3000," right click, and click ``Delete Rows 2529 - 3000." The only remaining rows are those with physical meaning.

\subsection{Plotting a Graph}
\begin{figure}[h]
    \centering
    \includegraphics[width=0.75\linewidth]{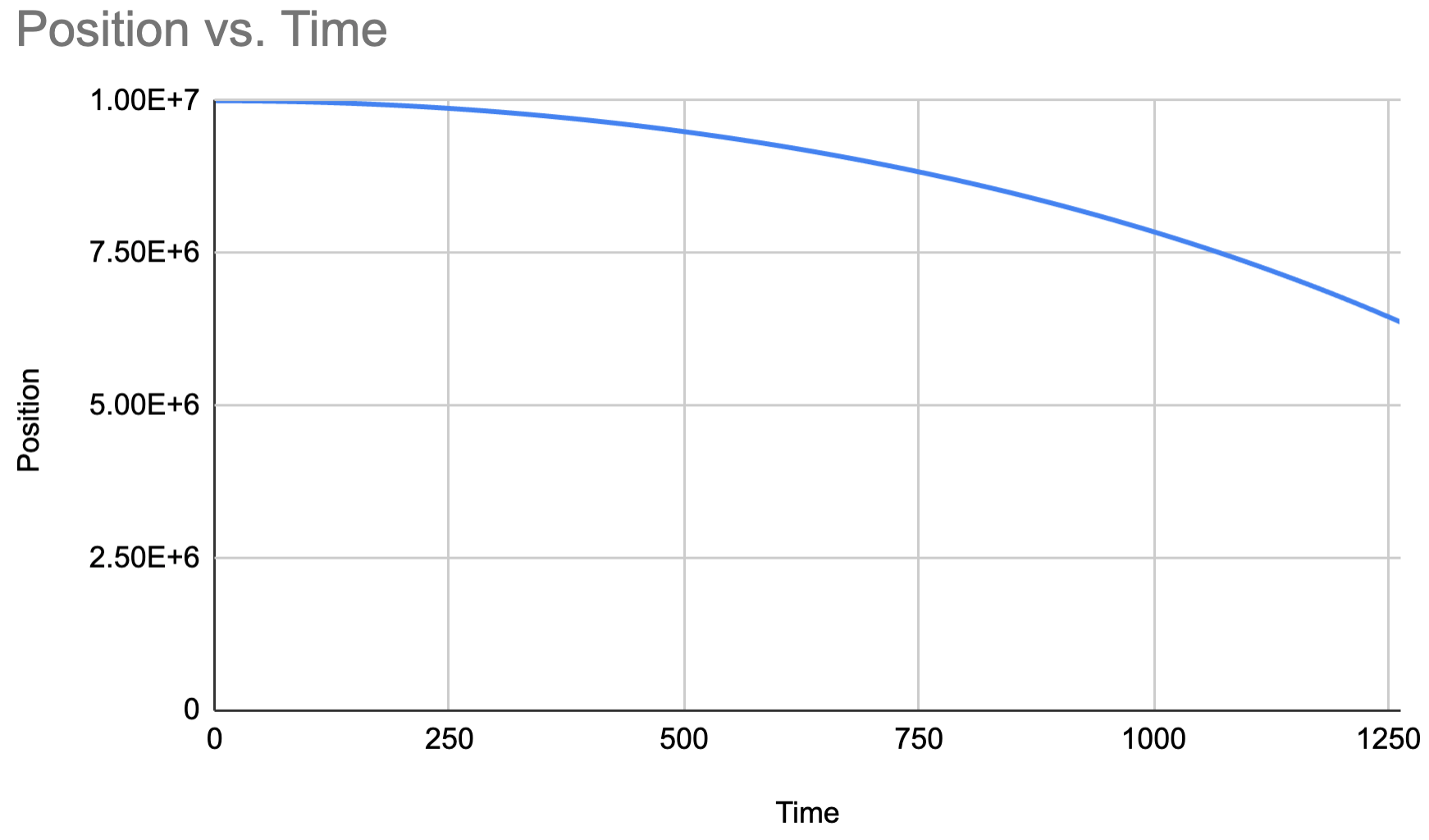}
    \caption{Google Sheets will usually visualize the motion of the falling astronaut with this line chart. Note that Figure \ref{numericalgraph.png} plots the same data, but the vertical axis begins at 6371000 m instead of 0 m. Figure \ref{numericalgraph.png} also includes the radius of the earth as a reference point. Students are encouraged to add units to the axes' labels for clarity.}
    \label{InstructionalChart.png}
\end{figure}

We can produce a graph of these data. Click the ``A" above the first column, then shift-click the ``B" to the right of A. Click the ``Insert" button from the toolbar, then click ``Chart." A chart will appear on the spreadsheet. Usually, an elegant position-versus-time graph will appear, one which resembles Figure \ref{InstructionalChart.png}.

If the graph does not look similar, there are a variety of troubleshooting techniques that can be done to fix the graph. Click on the top right corner of the graph and then click ``Edit Chart." (Note that the ``Chart Editor" menu may open automatically when inserting the graph.) Click on ``Chart Type" and set it to ``Line Chart" if it is not already. This may fix the issue.

If the graph still does not resemble Figure \ref{InstructionalChart.png}, look for the section labeled ``X-axis" and ensure that the data underneath states ``A2:A2528" if it does not already. Similarly, ensure that the data underneath the ``Series" section states ``B2:B2528." If the problem persists, the issue may be due to more subtle reasons than those listed here, and asking for help from the professor or Teaching Assistant may be required.

Note that the graph appears smooth and curved. In reality, the graph contains a variety of discrete points of data, each connected by a straight diagonal line. However, these data points are very close together compared to the size of the graph; each data point is 0.5 seconds apart from its neighbors, while the graph as a whole spans 1263 seconds. Thus, the straight diagonal lines are so short in length that the graph appears smooth and curved.

Students are encouraged to experiment with the features in the Chart Editor and the ``Customize" sub-menu. For example, upon clicking on the ``Vertical Axis" tab, students can adjust the minimum and maximum values that the graph displays on the position-axis. A student can set the minimum value to be 6371000 m, causing the graph to closer resemble Figure \ref{numericalgraph.png}. Note that students are encouraged to set a minimum value for 0 or greater, as Equation \ref{GoogleSheet} become un-physical for negative values of position. 

As an additional note, suppose a student wishes to have the graph appear at the top of the spreadsheet instead of the bottom. Such a location is useful if the student wishes to change the first few rows of the data and immediately see the graph update without scrolling down. To move the graph, click on it and copy it with CMD+C on a Mac or Ctrl+C on Windows. Delete it with ``delete" or ``backspace." Scroll to the top of the spreadsheet and paste the graph with CMD+V or Ctrl+V. This new location will be useful in the next section, in which we will change the data and immediately see its effects on the graph.

\subsection{Generalizing the Technique}

Suppose we wish to repeat this exercise but with different initial conditions. Perhaps we wish to model the motion of a falling space probe as it approaches a planet with a mass that is different from Earth's. Alternatively, we may wish to model the motion of a falling meteor which has some initial speed when it nears the Earth rather than being dropped from rest.

\subsubsection{Adjusting \textit{G} and \textit{M}}

Let us restructure the format of our spreadsheet so that we can quickly recalculate our position-time graph for any value of the universal gravitational constant $G$ or planet mass $M$. In Box C1, type ``Gravitational Constant." In Box C2, type ``=6.6743*10$^\wedge$(-11)." Upon entering this value, the box may display ``0" because the displayed values is rounding down due to $G$'s small size. However, when the software computes calculations with Box C2, the software will use the value for $G$ we gave it, not 0. In other words, the displayed value of 0 will not interfere with our calculations.

Next, type ``Planet Mass" into Box D1 and ``=5.97219*10$^\wedge$(24)" into Box D2. We have now explicitly recorded our values for $G$ and $M$ and have given them proper labels.

Let us now incorporate these values for $G$ and $M$ into our equation in Column B. Double-click on Box B4. Delete the numerical value representing $G$ and replace it with ``C\$2." The symbol ``\$" tells the computer that the equation is interested in that one particular box, no matter the iteration. As a point of comparison, when we took the equation as defined in Box B4 and inserted it into Box B5, the equation changed so that ``B2" became ``B3" and ``B3" became ``B4." In contrast, value ``C\$2" will remain ``C\$2."

Next, delete the numerical value representing $M$ and replace it with ``D\$2". Now that we have an updated equation, replace all of the the boxes in Column B rows 5 through 2528 with the new equation, using the technique described previously to quickly replace the values for thousands of cells.

The benefit of this new equation structure is that we can change the values for $G$ and $M$ using Boxes C2 and D2, respectively, and our position values will automatically recalculate, both in Column B and on the graph. Students are encouraged to experiment with different values of $G$ and $M$ and notice the result. In particular, try increasing the mass of the planet to ten times that of Earth or decreasing the mass to one-tenth that of Earth.

\subsubsection{Adjusting \texorpdfstring{\(\Delta t\)}{Δt}}

Suppose we want to make our approximation use a time-step other than 0.5 s. If we want to model a process that lasts for only 10 s, for example, then 0.5 s is no longer a small time-step compared to the whole process, and we will want our time-step value $\Delta t$ to be much smaller. Alternatively, we might realize that we can increase our time step without noticeably decreasing the accuracy of our results, and so we might increase our time step to shorten the number of iterations needed to reach the answer.

Type ``Time Step" into Box E1 and ``=0.5" into Box E2. In Box A3, replace ``0.5" with ``E\$2," and fill in the boxes beneath A3 with the same updated equation.

Similarly, double-click on Box B4 and replace ``0.5" with ``E\$2." Fill in the boxes beneath B4 with the updated equation.

We can now replace our time-step with whatever we desire by updating Box E2. Note that changing the time-step will not change the system itself -- for example, the astronaut will still take the same amount of time to reach the ground if we replace ``0.5" with ``5.0." However, our graph will change its appearance because upon changing the time-step, the amount of time spanned by 2528 iterations will change.

Students are encouraged to experiment with changing the time-step and observing its impact on the graph. If we change the time-step to 5.0 s, for example, the astronaut will reach the surface of the planet after just 255 iterations, making the spreadsheet much more succinct. However, the reported amount of time elapsed is slightly less accurate: the spreadsheet returns a value of 1265 s $< t <$ 1270 s, slightly greater than 1263 s. The greater time-step made the spreadsheet more efficient but less accurate. As a general principle, there exists a trade-off between time-steps that are small and precise and time-steps that are large and produce a manageable number of iterations.

\subsubsection{Adjusting The Initial Velocity}
\label{adjustingv0}

Suppose two bowling balls are released from the International Space Station. One ball is dropped gently with no initial speed, while the other is hurtled down with a large initial speed. The position-time graph of the second bowling ball will look quite different from the first. In this section, we explain how to adjust the initial conditions of the spreadsheet based on the initial velocity.

To determine how to implement an adjustable initial velocity, let us consider the following conundrum. In Equation \ref{key}, we see that each position $x_{n+1}$ is determined by the two positions immediately prior, $x_n$ and $x_{n-1}$. But consider the initial two positions $x_1$ and $x_0$, which do not have two immediately prior positions. How do we determine $x_0$ and $x_1$, considering that we cannot use Equation \ref{key}?

The initial position, $x_0$, must be given by the problem statement and cannot be deduced mathematically. Conceptually, this makes sense -- if we are asked to calculate the amount of time taken for the astronaut to fall to the ground, we would request to know from where the astronaut fell.

To find $x_1$, we must be given the initial velocity $v_0$. Conceptually, this also makes sense -- if we are asked where the astronaut is 0.5 seconds after leaving her spaceship, we would request to know how fast she was moving when she left the spaceship.

Suppose we are given $x_0$ and $v_0$. Based on Equation \ref{v_n-1}, we can infer that the initial velocity is:

\begin{equation}
\label{v0}
    v_0 \approx \frac{x_{1} - x_{0}}{\Delta t} \; .
\end{equation}

Our goal is to calculate $x_1$. Thus, we algebraically rearrange the equation to acquire:

\begin{equation}
\label{nextposition}
    x_{1}=v_0 \Delta t +x_0 \; .
\end{equation}

Note that the sign of the velocity matters. If a cannonball is shot down toward the ground from a spaceship, for example, it will arrive on the ground much sooner than if the cannonball had been shot upward away from the ground. Based on the coordinate system provided in Figure \ref{diagram1}, a negative velocity points toward the ground and a positive velocity points upward, away from the ground.

We can use this information to improve our spreadsheet. Type ``Initial Velocity" into Box F1, and enter your desired initial velocity into Box F2. Perhaps begin with ``=0" so that the value will not yet effect the position-time graph. Equation \ref{nextposition} tells us the astronaut's position immediately following the initial position based on the initial velocity. Change Box B3 (which represents $x_1$) to:

\begin{equation}
\label{nextpositionspreadsheet}
    \text{= F\$2 * E\$2 + B\$2} \;,
\end{equation}

which is Equation \ref{nextposition} rewritten for the spreadsheet.

We now have an efficient way to recalculate the motion of the astronaut with whatever initial velocity we desire.

Note that, in this particular case, we do not replace the boxes beneath B3 with the new equation. As stated earlier, the values $x_2$ and beyond are calculated with the two position values immediately prior. The values $x_1$ and $x_0$ were the exceptions because they were the only two values without two prior positions, and thus were the only two that could not be calculated with Equation \ref{key}. We were given the initial position directly from the problem statement, and thus Equation \ref{nextpositionspreadsheet} need be applied only to $x_1$ (Box B3).

Students are encouraged to experiment with different values for the initial velocity. In particular, when the time-step is 1.0 and the initial velocity is 3000, the motion of the object resembles that of a projectile, though this problem deals only with one spatial dimension.

\subsubsection{Adjusting the Initial Position}

To adjust the initial position, simply type a new value into Box B2.

The reader may ask: why was the method for adjusting the initial position reserved for the end of this section, considering that the method for adjusting this variable is far simpler than the other methods?

The reason why we reserved the initial position for last lies in the relationship between $x_0$, $x_1$, and $v_0$. Suppose the spreadsheet was arranged in the original setup we had prior to the updates described in this section. Now suppose we change the initial position (Box B2) from 10,000,000 m to 11,000,000 m. In our original setup, Box B3 was not defined in terms of Box B2 and the initial velocity, as it is in the updated version. Thus, changing Box B2 from 10,000,000 m to 11,000,000 m will leave box Box B3 unaffected, maintaining its value of 10,000,000 m. In doing so, we have implicitly told the computer that the astronaut travels 1,000,000 m in 0.5 seconds. This hardly seems like a plausible assumption!

In the original setup, initial velocity $v_0$ was not explicitly defined, but rather was implied by our choices for $x_0$ and $x_1$. In the new setup, $x_1$ is calculated by our choices for $x_0$ and $v_0$. Thus, the benefit of updating equation for $x_1$ prior to adjusting values for $x_0$ is that we can now change $x_0$ to whatever we desire without changing $v_0$. In other words, we input our desired initial velocity and initial position, and we do not accidentally tell the computer that the astronaut traveled a million meters in half a second.

Students are encouraged to experiment by adjusting all of the features discussed in this section: $G$, $M$, $\Delta t$, $v_0$, and $x_0$. What shapes can the student make with the position-time graph?

\subsection{Homework Assignment: A Large-Angle Pendulum}

This homework assignment is intended to be given to students after the 1 - 3 hour lab described above. Students are encouraged to work together as a group. They are intended to have one week to complete the assignment, which will provide time for students to form questions, identify mistakes, visit the professor during office hours, and compare their work with peers.

\subsubsection{Problem Statement for the Large-Angle Pendulum}

\begin{figure}[ht]
    \centering
    \includegraphics[width=0.35\linewidth]{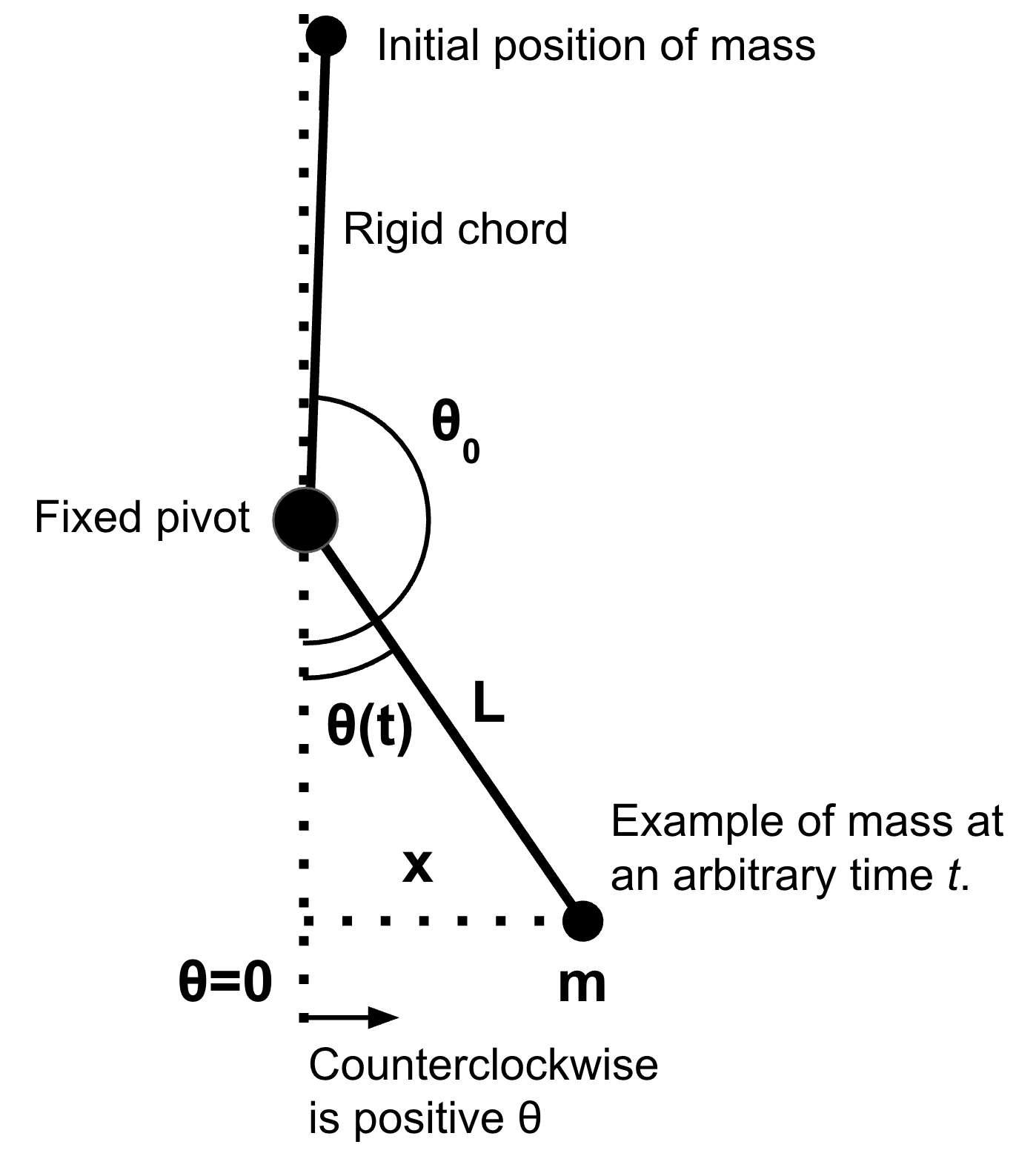}
    \caption{A diagram displaying the setup of the pendulum. The mass is released from an nearly vertical position, but slightly to the right. The chord is imagined to be inflexible, like that of a PVC pipe rather than a rope. Negative angles occur when the mass is to the left of the pivot.}
    \label{pendulum.png}
\end{figure}

We have a pendulum. The mass at the end of the chord can be modeled as a point particle, and the rigid chord can be approximated as massless. For example, suppose we super-glue a brick to one end of a meter stick, which has a mass much smaller than the brick, and we nail the other end of the meter stick to a rotating hinge on the wall. Now, we are free to pull the brick back some angle $\theta$ and let the brick swing back and forth, as shown in Figure \ref{pendulum.png}.

Note that $\theta$ = 0 corresponds to the mass being directly below the pivot, and $\theta$ = $\pi$ radians (180$^{\circ}$) corresponds to the mass being directly above the pivot.

The length of the chord is $L$, and the magnitude of the acceleration due to gravity is $g$ = 9.8 m s$^{-2}$.

The equation that describes the angular motion of this pendulum is:

\begin{equation}
\label{angular}
    \frac{d^2\theta}{dt^2} = - \frac{g}{L} \sin(\theta) \; .
\end{equation}

Suppose we pull back the mass to an angle of $\theta$ = 3.13, which is nearly vertical. We release the mass from rest and let it oscillate.\footnote{To help visualize this situation, visit the ``Simple Pendulum" computer simulation created by MyPhysicsLab.com: https://www.myphysicslab.com/pendulum/pendulum-en.html. Click on the swinging pendulum and drag it upwards until it is nearly vertical but ever-so-slightly tilted to the right. Credit: Erik Neumann, April 2001.}

Physicists are interested in modeling the motion of this pendulum over time. If we had pulled back the mass by some small angle, we could have approximated the position-versus-time motion of the mass with a sine or cosine function. However, this approximation is only accurate for small angles. Equation \ref{angular} is an inseparable differential equation, and thus we must evaluate it using a numerical approximation. 

\textit{(a)} Using the technique described in Section \Ref{SettingUp}, take Equation \ref{angular} (a continuous function) and transform it into a discrete equation. We will use this discrete equation to graph the motion of the pendulum in Part \textit{(c)} of this question. Your final answer should look like this:

\begin{equation}
\label{discretetheta}
    \theta_{n+1} = 2 \theta_n-\theta_{n-1}-\frac{g}{L}\sin(\theta_n) \Delta t^2\; .
\end{equation}

Be sure to show the step-by-step reasoning about how you arrived at this equation.

Hint: pay careful attention to Equation \ref{v_n}, Equation \ref{v_n-1}, and Equation \ref{discreteacceleration}.

\textit{(b)} We are told that the initial position is $\theta_0$ = 3.13 rad and that the initial speed is $v_0$ = 0 m s$^{-1}$. Using the technique described in Section \ref{adjustingv0}, explain why $\theta_1$ = 3.13 rad.

\textit{(c)} Use the technique described in Section \ref{Implementation} to create a digital spreadsheet modeling the motion of the mass. The spreadsheet should allow the user to change the values of $g$, $L$, $\theta_0$, $v_0$, and $\Delta t$, with the computer updating the spreadsheet upon changing the values. 

\begin{figure}[ht]
    \centering
    \includegraphics[width=0.75\linewidth]{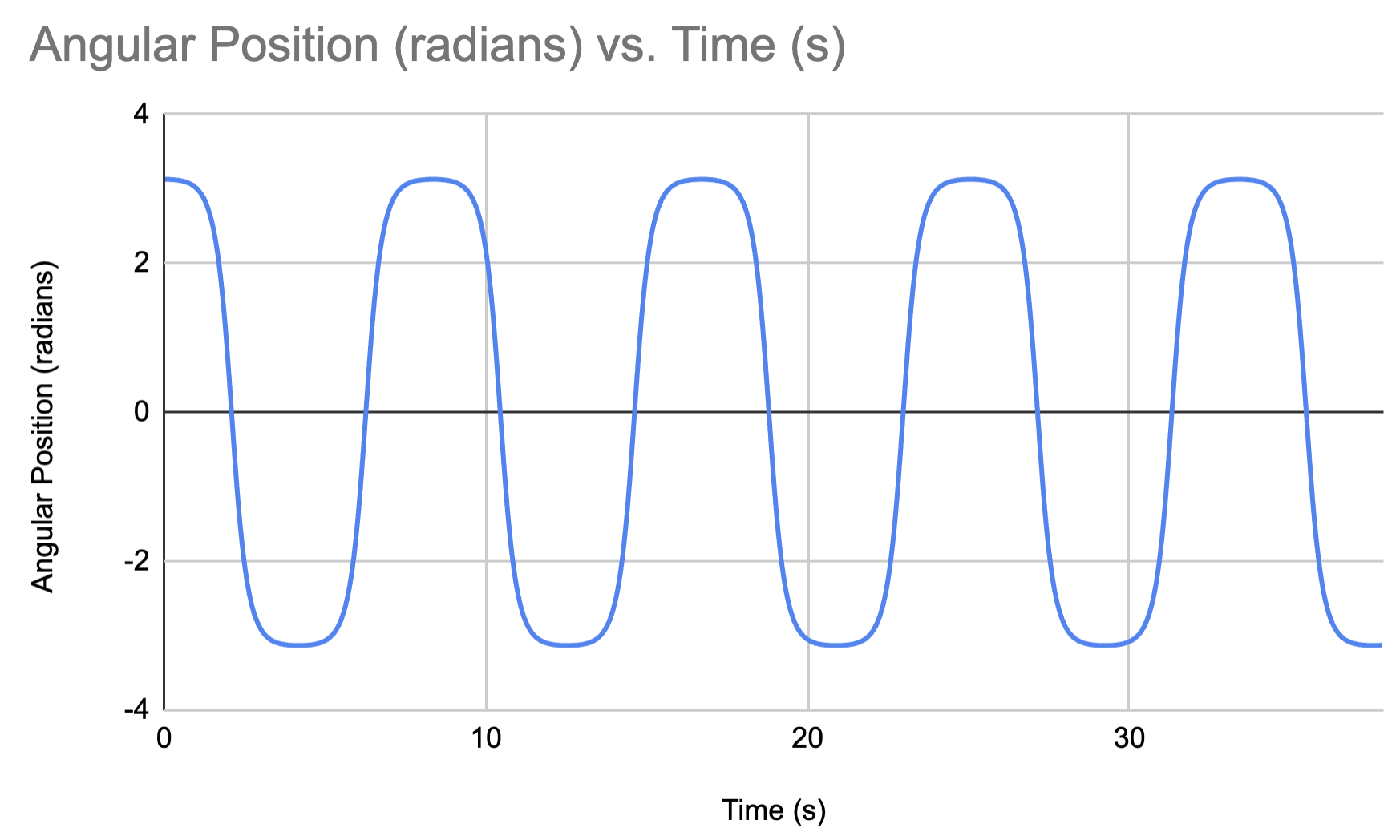}
    \caption{The angular position of the pendulum over time. Initially, the mass lingers near the top of its circular path, but it remains slightly to the right. After a moment, it swings down quickly, passes underneath the hinge, and continues upward until it is nearly at the zenith again but slightly to the left. It lingers there for a moment then reverses direction.}
    \label{fig:LargePendulum}
\end{figure}

\textit{(d)} Include a graph plotting angular position over time. Your graph should resemble Figure \ref{fig:LargePendulum}.

Tip 1: Optionally, you may begin your new spreadsheet by opening the spreadsheet for the Falling Astronaut Problem and clicking on ``Files" then ``Make a Copy" to duplicate your presentation. Thus, you can create your spreadsheet for the large-angle pendulum by modifying an existing spreadsheet rather than beginning from scratch.

Tip 2: The amount of time a one-meter-long pendulum takes to make one full oscillation is probably much less than 1263 seconds, and so our time step should be much smaller than 0.5 seconds. Try experimenting with different values, such as 0.01 s.

Tip 3: The vertical axis is measuring angular position, and so the values will be within the range $- 2 \pi \le  \theta \le 2 \pi$. Pay careful attention to the minimum and maximum values on the vertical axis that have been set, if any, as the vertical axis values in this new graph will be much less than 6371000.

Tip 4: Remember that $\theta_2$ and beyond will be defined by Equation \ref{angular}, $\theta_0$ is typed in by hand, and $\theta_1$ is defined by the technique described in Section \ref{adjustingv0}.

\textit{(e)} Extra-credit question: Figure \ref{fig:LargePendulum} models angular position, but what about the horizontal position $x(t)$? The angular position $\theta(t)$ and horizontal position $x(t)$ are related by:

\begin{equation}
    \begin{gathered}
     \sin \big( \theta(t) \big) = \frac{x(t)}{L}. \\[10pt]    
    L \sin \big( \theta(t) \big) = x(t).
    \end{gathered}
\end{equation}

\begin{figure}[ht]
    \centering
    \includegraphics[width=0.75\linewidth]{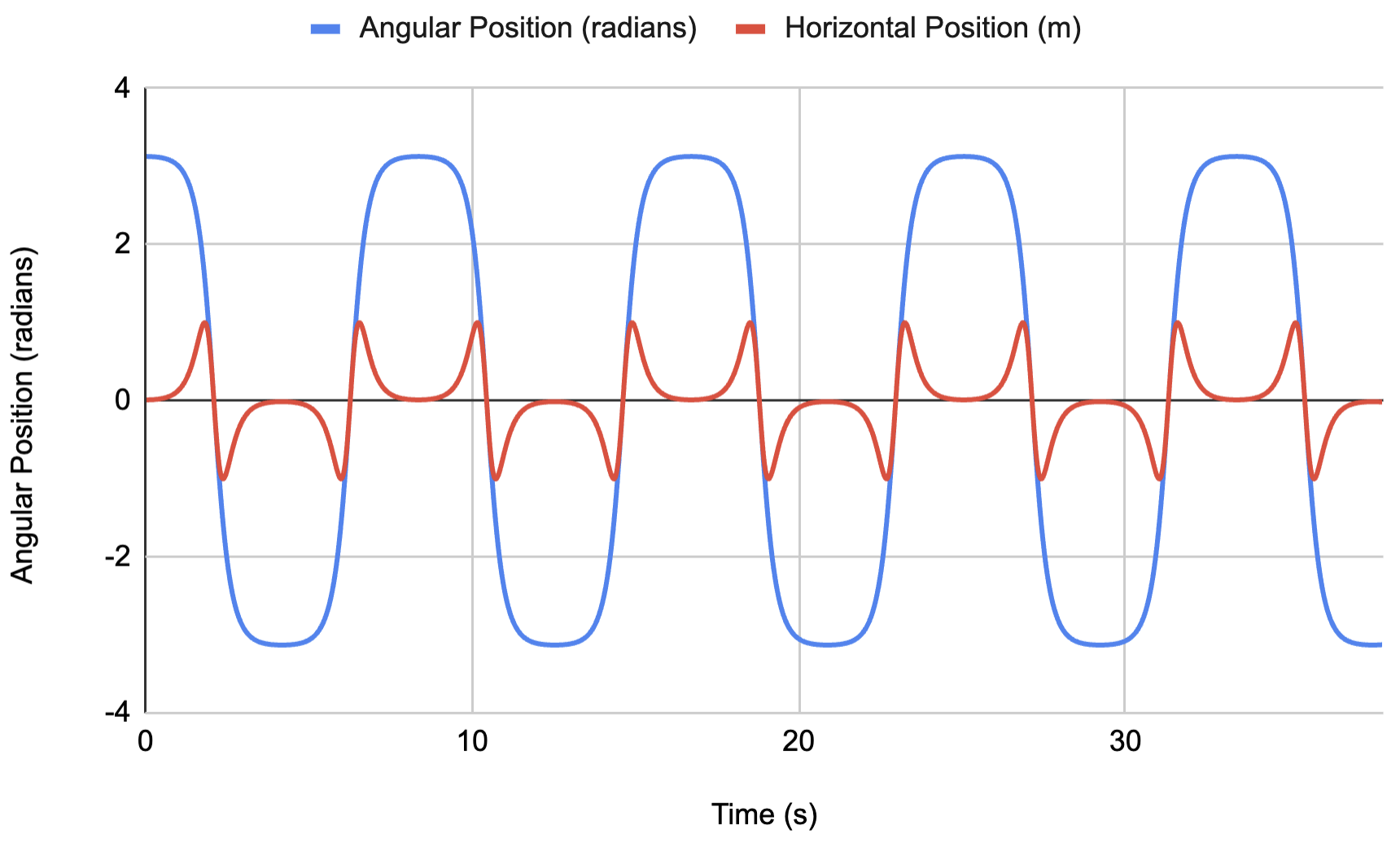}
    \caption{A comparison of the angular position in blue and the linear horizontal position in red. The pivot is defined to be at $x$ = 0 m, with positive values of $x$ being to the right of the pivot. When the mass lingers near $\theta= \pi$, the position is nearly zero, as the mass is almost directly above the pivot. When the mass zooms beneath the pivot, the horizontal position is also $x$ = 0, but only for a brief moment as the speed of the mass is large.}
    \label{PositionandAngle.png}
\end{figure}

Using this equation, create another column in the spreadsheet, one which models the horizontal position $x$ over time. When the angular position and horizontal position are plotted together, the resulting graph should resemble Figure \ref{PositionandAngle.png}.

\subsection{Answer Key to Homework Assignment}

\subsubsection{Answer to Part (a)}

 We are interested in discretizing $\frac{d^2 \theta}{dt^2}$, the angular acceleration, which can also be written as $\alpha$. Our strategy will be to take the method for discretizing the linear acceleration, $a$, and converting from linear kinematics to rotational kinematics. Similarly, we will take equations for the linear velocity $v$ and convert them to angular velocity $\omega$. With regard to discretizing the linear acceleration, let us inspect Equation \ref{discreteacceleration}, which states that 

\begin{equation}
    a_n = \frac{v_n-v_{n-1}}{\Delta t} .
\end{equation}

Note that this equation uses equality symbol, however the equation is technically an approximation. In the context of the numerical approximation, the two values are close enough that we can treat the equation as if it is an equality.

This is a discrete equation for the acceleration, which is what we want. Let us now convert it to angular kinematics. Angular acceleration and linear acceleration are related to each other by the radius of the circular path the rotating object takes, $r$, as follows:

\begin{equation}
   a  = r \alpha \Rightarrow \frac{a}{r} = \alpha.
\end{equation}

Similarly, angular velocity and linear velocity are related with the following equation:

\begin{equation}
   v  = r \omega \Rightarrow \frac{v}{r} = \omega.
\end{equation}

Thus, to convert our equation from a linear equation to a rotational equation, we simply divide by the radius of the circular path:

\begin{equation}
\label{alphaalpha}
    \begin{gathered}
    \frac{a_n}{r} = \frac{1}{r} \left( \frac{v_n-v_{n-1}}{ \Delta t} \right) = \frac{v_n/r-v_{n-1}/r}{ \Delta t} \\[10pt]
    \Rightarrow \alpha_n = \frac{\omega_n-\omega_{n-1}}{\Delta t} .  
    \end{gathered}
\end{equation}

We have discretized the angular acceleration, but there is still a problem. Our ultimate goal is to have an equation in which we input the current angular position $\theta_n$ and the immediately previous angular position $\theta_{n-1}$ to determine the next position $\theta_{n+1}$. Equation \ref{alphaalpha} is written in terms of angular velocity $\omega$, not angular position. To switch from angular velocity to angular position, we will convert Equation \ref{v_n} and Equation \ref{v_n-1} from linear kinematics to rotational kinematics:

\begin{equation}
\label{omegan}
    \begin{gathered}
    v_n=\frac{x_{n+1}-x_n}{\Delta t} \\[10pt]
    \Rightarrow \frac{v_n}{r} = \frac{1}{r} \left( \frac{x_{n+1}-x_{n}}{ \Delta t} \right) = \frac{x_{n+1}/r-v_{n}/r}{ \Delta t} \\[10pt]
    \Rightarrow \omega_n = \frac{\theta_{n+1}-\theta_{n}}{\Delta t}
    \end{gathered}
\end{equation}

and

\begin{equation}
\label{omegan-1}
    \begin{gathered}
    v_{n-1}=\frac{x_n-x_{n-1}}{\Delta t} \\[10pt]
    \Rightarrow \frac{v_{n-1}}{r} = \frac{1}{r} \left( \frac{x_{n}-x_{n-1}}{ \Delta t} \right) = \frac{x_{n}/r-v_{n-1}/r}{ \Delta t} \\[10pt]
    \Rightarrow \omega_{n-1} = \frac{\theta_{n}-\theta_{n-1}}{\Delta t} .  
    \end{gathered}
\end{equation}

Note that in Equations \ref{alphaalpha}, \ref{omegan}, and \ref{omegan-1}, the angular versions of the equations have an extremely similar structure to their linear counterparts -- we simply replaced $a$ with $\alpha$, $v$ with $\omega$, $x$ with $\theta$.

Now that we have clearly defined angular velocities, we can substitute Equation \ref{omegan} and Equation \ref{omegan-1} into Equation \ref{alphaalpha} to yield:

\begin{equation}
\label{alphaalphalpha}
    \alpha= \frac{\omega_{n} - \omega_{n-1}}{\Delta t}= \left( \frac{\theta_{n+1} - \theta_n}{\Delta t} - \frac{\theta_n - \theta_{n-1}}{\Delta t} \right) \frac{1}{\Delta t} = \frac{\theta_{n+1} - 2\theta_n + \theta_{n-1}}{\Delta t^2}.
\end{equation}

We have successfully discretized the angular acceleration ($\alpha=\frac{d^2\theta}{dt^2}$) in terms of angular position. Our original goal was to discretize the equation of motion for the pendulum, Equation \ref{angular}. Let us substitute our equation for the discretized angular acceleration, Equation \ref{alphaalphalpha}, into Equation \ref{angular}. We will also replace the continuous value $\theta$ with the discrete value $\theta_n$:

\begin{equation}
    \frac{\theta_{n+1} - 2\theta_n + \theta_{n-1}}{\Delta t^2} = - \frac{g}{L}\sin(\theta_n) .
\end{equation}

Algebraically rearranging this equation yields:

\begin{equation}
    \boxed{
    \theta_{n+1} = 2\theta_n- \theta_{n-1}-\frac{g}{L}\sin(\theta_n) \Delta t^2 . }
\end{equation}

This equation matches Equation \ref{discretetheta} exactly.

As reminder, the physical meaning behind this equation is as follows: if we know the position of the pendulum at some instant of time and the position immediately prior to that instant, we can determine the position at the next instant in time. We can use this new position value to calculate the position at one instant of time after that, and the instant after that, and so on and so forth.

\subsubsection{Answer to Part (b)}

The goal of this problem is to find the position of the mass immediately one time-step after being released. Mathematically, this value is $\theta_1$. Let us inspect Equation \ref{nextposition}, which we used to find $x_1$:

\begin{equation}
    x_{1}=v_0 \Delta t +x_0 \; .
\end{equation}

We can convert this equation from linear kinematics to rotational kinematics by dividing by the radius of the circular path $r$, as we did in the previous section:

\begin{equation}
    \begin{gathered}
    \frac{x_{1}}{r}= \frac{v_0 \Delta t}{r} + \frac{x_0}{r} \; . \\[10pt]
    \Rightarrow\theta_{1}=\omega_0 \Delta t +\theta_0 \; . 
    \end{gathered}
\end{equation}

We are told that the initial position is $\theta_0=$ 3.13 rad, and that the mass is released from rest ($\omega_0=0$ rad s$^{-1}$). Substituting these values, we acquire 

\begin{equation}
\label{nextpositionangular}
    \begin{gathered}
    \theta_{1}= 0 \frac{\text{rad}}{\text{s}} \times \Delta t + 3.13 \; \text{rad} \\[10pt]
    \Rightarrow \boxed{\theta_{1}= 3.13 \; \text{rad},}
    \end{gathered}
\end{equation}

which matches the value provided in the problem statement. Note that we could also make a conceptual argument for why $\theta_1=$ 3.13 rad: the mass begins at a position of 3.13 rad, and the object is initially not moving, so at the end of the first time step, there has been no change in its position -- it is still located at 3.13 rad.

\subsubsection{Answer to Parts (c), (d), and (e)}

The answers to Parts $\textit{(c)}$, $\textit{(d)}$, and $\textit{(e)}$ are presented in a full excel spreadsheet in the file \\
$\texttt{Answer\_Key\_Large-Angle\_Pendulum.xlsx}$, which is available in the source files in the arXiv submission for this paper. Readers can download the source package to access this file.

\section{Concluding Remarks}

This paper has attempted to show that numerical methods for evaluating differential equations do not require years of technical experience, as some undergraduate curricula implicitly assume, but rather that first-year undergraduate students can gain a basic level of understanding of the technique in an afternoon and can strengthen their comfort using the technique in a week.

Although the technique requires introducing students to discretization, iterating on equations, and use of digital data spreadsheets, the underlying mathematics requires only the definitions of the first and second derivatives and algebraic manipulation.

The techniques presented in this paper bring a multitude of otherwise advanced physics problems within the reach of undergraduate students. Chaotic systems like the three-body problem or double-pendulum, for example, possess no analytical solution and require numerical approximations. Numerical approximations also appear in astronomy and General Relativity, including the precession of the perihelion of Mercury's orbit. With techniques such as the one described in this paper, students can confidently navigate questions which are usually considered beyond the scope of an introductory college physics class.

This paper aims to provide students with a new style of thinking and a deeper appreciation for modeling motion through differential equations. Some physics textbooks implicitly assume that analytical solutions are the more elegant format for presenting an equation, while differential equations are treated as an unsatisfying last resort when no analytical solutions are available. The aim of the Falling Astronaut Problem is to demonstrate that, in some cases, numerical methods for evaluating a differential equation are more efficient and elegant than the analytical form. The author hopes that the laboratory exercise and homework assignment provided in Section \ref{Implementation} will help students conceptualize cause-and-effect through a new lens, and in doing so, gain a new tool for describing motion in the universe.

\section{Acknowledgments}

Thank you to Laura Fledderman for her advice and support during the early drafts of this paper.

\end{document}